\documentclass[12pt]{article}
\usepackage{amsmath,amssymb}
\usepackage[hypertex]{hyperref}
\usepackage{graphicx}
\usepackage{color}
\usepackage{here}

\numberwithin{equation}{section}

\def\bb#1{\mathbb{#1}}
\def\({\left(}
\def\){\right)}
\def\pare#1{\left( #1\right)}
\def\bpare#1{\left\{ #1\right\}}
\def\cpare#1{\left[ #1\right]}

\def\abs#1{\left| #1\right|}
\def\defeq{\ \raisebox{0.2pt}{:}\hspace{-1.2mm}=}
\def\ds{\displaystyle}
\def\ssp{\hspace{0.3mm}}
\def\am{\mathrm{am}\hspace{0.2mm}}
\def\sn{\,\mathrm{sn}}
\def\cn{\,\mathrm{cn}}
\def\dn{\,\mathrm{dn}}
\def\iom{i\hspace{0.2mm}\omega}
\def\iomm#1{i\hspace{0.2mm}\omega_{#1}}

\def\eK{\mathrm{\bf K}}
\def\eE{\mathrm{\bf E}}
\def\eF{\mathrm{\bf F}}

\def\RS#1{$\mathbb{R}_{\rm t} \! \times {\rm S}^{#1}$}

\def\AdSxS{{AdS${}_5 \times {}$S${}^5$}}
\renewcommand{\eqref}[1]{$\pare{\rm \ref{#1}}$}

\def\cQ{{\mathcal Q}}
\def\cO{{\mathcal O}}

\DeclareMathOperator{\arcsinh}{arcsinh}

\newcommand{\bmt}[1]{{{\mbox{\boldmath$ #1 $}}}}

\begin{document}

%%%%%%%%%%%%%%%%%%%%%%%%%
{\ }
\vspace{-10mm}

\begin{flushright}
{\bf January 2008}\\[1mm]
{\small UT\,-\,07\,-\,41}
\end{flushright}

\vskip 2cm

\begin{center}
\LARGE
%%%%%%%%%%%%%%%%%%%%%%%%

\mbox{\bf Finite\,-\ssp Size Effects for Dyonic Giant Magnons}

%%%%%%%%%%%%%%%%%%%%%%%%

\vskip 2cm
\renewcommand{\thefootnote}{$\alph{footnote}$}

\large
\centerline{\sc
Yasuyuki Hatsuda\,\footnote{{\tt \,hatsuda@hep-th.phys.s.u-tokyo.ac.jp}}
\quad and \quad\ 
Ryo Suzuki\,\footnote{{\tt \,ryo@hep-th.phys.s.u-tokyo.ac.jp}}}

\vskip 1cm

\emph{Department of Physics, Faculty of Science, University of Tokyo,\\
Bunkyo-ku, Tokyo 113-0033, Japan}

\end{center}

%%%%%%%%%%%

\vskip 14mm

\centerline{\bf Abstract}

\vskip 6mm

We compute finite-size corrections to dyonic giant magnons in two ways. One is by examining the asymptotic behavior of helical strings of {\tt hep-th/0609026} as elliptic modulus $k$ goes to unity, and the other is by applying the generalized L\"{u}scher formula for $\mu$-term of {\tt arXiv:0708.2208} to the situation in which incoming particles are boundstates.
By careful choice of poles in the $su(2|2)^2$-invariant $S$-matrix, we find agreement of the two results, which makes possible to predict the (leading) finite-size correction for dyonic giant magnons to all orders in the 't Hooft coupling.

\vspace*{1.0cm}

\vfill
\thispagestyle{empty}
\setcounter{page}{0}
\setcounter{footnote}{0}
\renewcommand{\thefootnote}{\arabic{footnote}}
\newpage

%%%%%%%%%%%%%%%%%%%%%%
%%%%%%%%%%%%%%%%%%%%%%
\section{Introduction}

There has been great advance toward understanding the correspondence between ${\cal N}=4$ super Yang-Mills and superstring on \AdSxS\ recently. The AdS/CFT correspondence \cite{Maldacena97, GKP98, Witten98} predicts a map between individual string states and gauge invariant operators in super Yang-Mills at least for large $N$, and under this map energy of a string state should be equal to conformal dimension of the corresponding operator.

Progress on checking this correspondence has been catalyzed by the discovery of integrability. The dilatation operator of ${\cal N}=4$ super Yang-Mills theory is shown to have the same form as Hamiltonian of an integrable spin chain, which enables us to study the problem of diagonalizing the Hamiltonian by a technique called Bethe Ansatz \cite{MZ02}. Long-range Bethe Ansatz equations for the full $psu(2,2|4)$ sector are proposed to an arbitrary order of the 't~Hooft coupling $\lambda \equiv N \ssp g_{\rm YM}^2$ in \cite{BS05}, assuming all order integrability of super Yang-Mills theory. Their original proposal contained so-called dressing phase, which was first introduced in \cite{AFS04}. The dressing phase reconciles mismatch between scaling limit of the Bethe Ansatz equations of \cite{BDS04} and the integral equation derived from classical string theory \cite{KMMZ04}. An all-order expression of the dressing phase was later proposed in \cite{BHL06, BES06} on the assumptions of transcendentality \cite{KL01, KL02, KLOV04} and crossing symmetry \cite{Janik06, AF06}.

However, the long-range Bethe Ansatz equations equipped with the dressing phase can reproduce the correct answer of super Yang-Mills only when the length of spin chain $L$ is large enough. For spin chains with finite size, the Bethe Ansatz equations do not account for wrapping interactions \cite{BDS04}, which possibly arise from the order of $\lambda^{L}$ as higher-genus diagrams \cite{ST05}.
In fact, the Bethe Ansatz prediction is found to disagree with the BFKL prediction \cite{Lipatov76, KLF77, Lipatov78} in \cite{KLRSV07}.
Recently, it is found that the wrapping effects for the four-loop anomalous dimensions of certain short operators induce terms of higher degrees of transcendentality \cite{FSSZ07, KM08}.

The wrapping problem does not occur for the system of infinite $L$ at weak coupling, and such situation has been studied under the name of asymptotic spin chain \cite{Staudacher04, Beisert05}.
It was shown that the $S$-matrix of the asymptotic spin chain can be determined only by the symmetry algebra $psu(2|2)^2 \ltimes {\bb R}^3$ up to the dressing phase, and that its BPS relation constrains the dispersion relation of magnon excitations as $\varepsilon (p) = \cpare{1 + f(\lambda) \sin^2 \pare{ \frac{p}{2}} }^{1/2}$, where the function $f(\lambda)$ is conjectured as the one given in \eqref{nonrel dispersion}.
On string theory side, the asymptotic spin chain corresponds to the states with an infinite angular momentum. Classical string solutions which correspond to elementary magnon excitations over the asymptotic spin chain are found in \cite{HM06} and called giant magnons. This correspondence is subsequently generalized to the one between magnon boundstates \cite{Dorey06} and dyonic giant magnons \cite{CDO06a}.

The $S$-matrix of classical worldsheet theory on \AdSxS\ was examined in \cite{KMRZ06}. It was further found that the $psu(2|2)^2 \ltimes {\bb R}^3$ symmetry is realized in the worldsheet $S$-matrix when the level matching conditions are relaxed \cite{AFPZ06, AFZ06b}.
Moreover in \cite{AFZ06b}, they proposed ``string" $S$-matrix which satisfies the standard Yang-Baxter equation, while ``gauge" $S$-matrix of \cite{Beisert05} satisfies the twisted Yang-Baxter equation.

With remarkable success for the case of infinite $L$ in mind, a natural question is what will be the dispersion relation of asymptotic spin chain when $L$ is finite. From string theoretical point of view, answering to this question boils down to construction of classical string solutions with finite angular momenta which incorporate (dyonic) giant magnons.\footnote{In conformal gauge, the ``size" can be interpreted also as the circumference of worldsheet.} Such solutions have already been constructed; see \cite{AFZ06a, AFGS07} for a general solution including giant magnons, and \cite{OS06} for solutions including dyonic giant magnons. And it has been found in \cite{AFZ06a} that the energy-spin relation of giant magnons receives correction of the order $e^{- c J_1}$, with $J_1$ the angular momentum along a great circle of ${\rm S}^2 \subset {\rm S}^5$ and $c = 2 \pi / [\sqrt \lambda \sin \pare{\frac{p}{2}} ]$.

\begin{figure}[t]
\begin{minipage}[t]{.47\textwidth}
\centering
\includegraphics[scale=0.8]{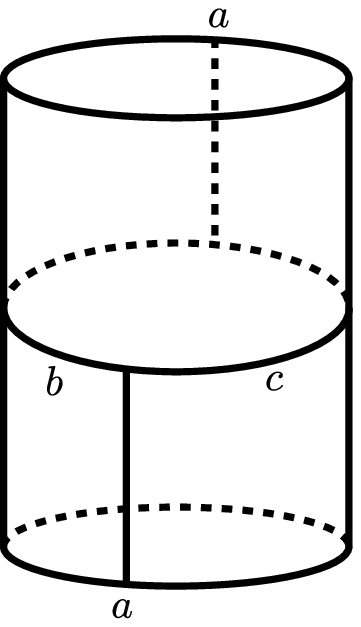}
\end{minipage}
\hfill
\begin{minipage}[t]{.47\textwidth}
\centering
\includegraphics[scale=0.8]{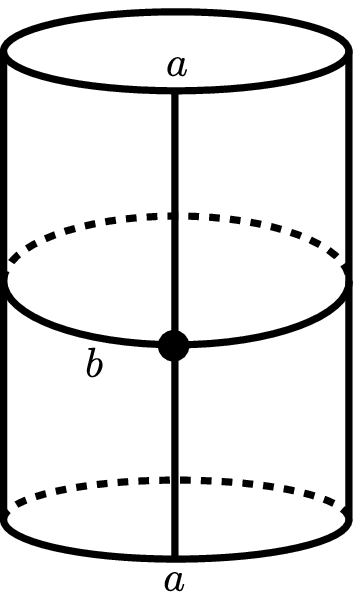}
\end{minipage}
\caption{Diagrams for the leading finite-size corrections. The left is called $\mu$-term, and the right $F$-term.
$a$ is an incoming physical particle, and $b,\,c$ are virtual (but on-shell) particles.}
\label{fig:diagram}
\end{figure}

It is argued in \cite{AJK05} that the exponential finite-size correction at strong coupling is related to the wrapping interaction at weak coupling, based on Thermodynamic Bethe Ansatz approach \cite{Zamolodchikov90, DT96, DT97} and the L\"uscher formula \cite{Luscher83, Luscher86, KM91}.
Recently, Janik and \L ukowski have elaborated this argument \cite{JL07}, assuming that L\"uscher's argument can be applied to the non-relativistic dispersion relation
\begin{equation}
\varepsilon (p) = \sqrt{1 + \frac{\lambda}{\pi^2} \sin^2 \pare{ \frac{p}{2}} } \,.
\label{nonrel dispersion}
\end{equation}
Their ``generalized L\"uscher formula" computes finite\,-$J_1$ correction to the energy-spin relation of giant magnons from the $S$-matrix and the dispersion relation \eqref{nonrel dispersion} of infinite\,-$J_1$ system. Since we know the conjectured $S$-matrix and dispersion relation of infinite-size system, the generalized L\"uscher formula will in principle give the finite-size correction valid at arbitrary values of $\lambda$. However, just like the original L\"uscher formula, it is only sensitive to the leading part of corrections exponentially suppressed in $L$ (or $J_1$), that is the first term in the following expansion:
\begin{equation}
\delta \varepsilon (p) = \alpha (p, \lambda, L) \, e^{- c (p, \lambda) L} + {\cal O} \, ( e^{-c' (p, \lambda) L} ) \qquad {\rm with} \quad c' (p, \lambda) > c (p, \lambda),
\label{def:leading corr}
\end{equation}
where $\alpha (p, \lambda, L)$ contains no factor exponentially dependent on $L$. According to the (generalized) L\"uscher formula, the leading finite-size correction arises from exchanging virtual particles going around the worldsheet cylinder once, and is written as
\begin{equation}
\delta \varepsilon (p) = \delta \varepsilon^\mu (p) + \delta \varepsilon^F (p) \,.
\label{leading correction Fmu}
\end{equation}
The first term is called $\mu$-term and the second one is called $F$-term, which have different diagrammatic interpretation as shown in Figure \ref{fig:diagram}.

Janik and \L ukowski computed the $\mu$-term of their generalized formula and found, after taking contributions from the BHL/BES dressing phase \cite{BHL06, BES06} into account, that
\begin{equation}
\alpha (p, \lambda, L) \, e^{-c L} \Big|_{\mu {\rm -term}} \approx - \frac{4 \sqrt \lambda}{\pi} \, \sin^3 \pare{\frac{p}{2}} \exp \cpare{- \frac{2 \pi L}{\sqrt \lambda \sin \pare{\frac{p}{2}}} - 2} \quad \pare{{\rm as} \ \lambda, L \to \infty},
\label{leading correction mu}
\end{equation}
which correctly reproduces the leading finite\,-$J_1$ correction to the dispersion relation of giant magnons in conformal gauge, with $L = J_1$ \cite{AFZ06a, AFGS07}.\footnote{What corresponds to the $F$-term in string theory, is not discussed in \cite{JL07}. Indeed, the exponential part of $F$-term seems to be different from that of $\mu$-term, so we do not discuss $F$-term in the main text.}

In this paper, we extend their analysis and study the leading finite-size correction to magnon boundstates and dyonic giant magnons. Firstly, we analyze the asymptotic behavior of helical strings of \cite{OS06} in the limit when they nearly reduce to an array of dyonic giant magnons, and determined the leading finite\,-$J_1$ correction to the energy-spin relation. Secondly, we apply the generalized L\"uscher formula for $\mu$-term to the situation in which the incoming particle is magnon boundstate.

Since the generalized L\"uscher formula of Janik and \L ukowski is applicable only to incoming elementary magnons, we slightly generalize their argument, assuming there exists an effective field theory such that it reproduces the non-relativistic dispersion
\begin{equation}
\varepsilon_Q (p) = \sqrt{Q^2 + \frac{\lambda}{\pi^2} \sin^2 \pare{ \frac{p}{2}} } \,,
\label{nonrel dispersion Q-mag}
\end{equation}
and the $S$-matrix which is given by the product of the conjectured two-body $S$-matrices.
Our results serve as a consistency check between the generalized L\"uscher formula and the results from string theory. It is desirable if one can give further justification of these formulae from other methods of computing the finite-size corrections.

Also we would like to stress that evaluation of the formula is not straightforward. Evaluation of the $\mu$-term requires information of residue at the poles that are located at the nearest from the real axis. Thus, to compute the $\mu$-term correctly, we have to determine which poles of the $su(2|2)^2\ S$-matrix are relevant.

Firstly we solve the condition of energy-momentum conservation associated to the splitting process shown on the left of Figure \ref{fig:diagram}. Secondly, we look for the poles of $S$-matrix that are consistent with the energy-momentum conservation. The diagrammatic technique developed in \cite{DHM07, DO07} turns out to be useful for this purpose, where they showed how simple and double poles of the $su(2|2)^2\ S$-matrix are related to particular exchange of physical particles.

After evaluating the residues of all such poles, we compare the result of the L\"uscher formula with that of classical string. By suitably modifying the contour of integration, we find the two results agree for both of the $Q \sim \cO( \lambda^{1/2} ) \gg 1$ and $Q \sim \cO(1) \ll \lambda^{1/2}$ cases.

\medskip
The paper is organized as follows.
In Section \ref{sec:DGM}, we discuss finite\,-$J$ correction to dyonic giant magnon from classical string solutions.
In Section \ref{sec:boundstate}, we apply the generalized L\"uscher formula for $\mu$-term to the cases in which the incoming particle is a magnon boundstate.
Section \ref{sec:discussion} is devoted to the comparison of the two results, discussion and conclusion.
In Appendices \ref{sec:definition}, \ref{sec:expansion}, \ref{sec:detail}, we collect the details of calculation needed to derive the results in the main text.
In Appendix \ref{app:Luscher}, we briefly review derivation of the generalized L\"uscher formula, slightly modifying the argument of \cite{JL07} to the case of our interest.
We give brief discussion on $F$-term in Appendix \ref{app:F-term}.

%%%%%%%%%%%%%%%%%%%%%%
\section{Finite\,-$\bmt{J}$ Correction to Dyonic Giant Magnons}\label{sec:DGM}

Dyonic giant magnon is a classical string solution on \RS{3} \cite{CDO06a}, which is two-spin generalization of the giant magnon solution found in \cite{HM06}. It has one infinite angular momentum $J_1$ and another finite angular momentum $J_2$\,, and obeys the square-root type energy-spin relation:
\begin{equation}
E - J_1 = \sqrt{ J_2^2 + \frac{\lambda}{\pi^2} \, \sin^2 \pare{ \frac{p_1}{2} } } \,.
\label{energy-spin dgm}
\end{equation}
The finite\,-$J_1$ generalization of dyonic giant magnon is found in \cite{OS06}, and named ``helical string" after the helical-wave solution of Complex sine-Gordon system. Thus, conserved charges and winding numbers of the helical string provide us with sufficient information about the finite\,-$J_1$ correction to the energy-spin relation \eqref{energy-spin dgm},

We use the notation of \cite{OS06} throughout this section.

\subsection{Dyonic giant magnon}

We begin with the review on $J_1=\infty$ case: the dyonic giant magnon. Dyonic giant magnon can be obtained by taking $k$, the elliptic modulus of helical strings, to unity.

\bigskip
The profile of dyonic giant magnons can be written as
\begin{gather}
t = a T + b X  \,, \qquad
\xi_{1} = \frac{\sinh(X  - \iomm{1})}{\cosh(X )}
\, e^{i \tan(\omega_1)X + i u_1 T}\,,\qquad 
\xi_{2} = \frac{\cos \omega_1}{\cosh(X )} \;
\, e^{i u_2 T} \,,
\label{profile dgm} \\[1mm]
T (\tau,\sigma) \equiv \frac{\tilde \tau - v \ssp \tilde \sigma}{\sqrt{1 - v^2}} \,,\qquad
X (\tau,\sigma) \equiv \frac{\tilde \sigma - v \ssp \tilde \tau}{\sqrt{1 - v^2}} \,,\qquad
\pare{\tilde \tau, \tilde \sigma} \equiv \pare{\mu \tau, \mu \sigma},
\label{def:X,T}
\end{gather}
where $a \,, b \,, v \,, u_1$ are parameters determined by $\omega_1$ and $u_2$\,. The parameter $\mu$ determines periodicity as $\tilde \sigma \simeq \tilde \sigma + 2 \pi \mu$. For dyonic giant magnons we set $\mu = \infty$, which relaxes the periodicity condition for a closed string in spacetime.

The conserved charges for a single dyonic giant magnon, rescaled by $\pi/\sqrt \lambda$, are given by
\begin{equation}
\begin{array}{ccccl}
{\cal E} &\defeq &\ds \frac{\pi}{\sqrt \lambda} \, E &= &\ds {u_1 \pare{1 - \frac{\tan^2 \omega_1}{u_1^2}} \eK (1)} \,, \\[5mm]
{\cal J}_{1} &\defeq &\ds \frac{\pi}{\sqrt \lambda} \, J_{1} &= &\ds {u_1 \cpare{ \pare{1 - \frac{\tan^2 \omega_1}{u_1^2}} \eK (1) - \cos^2 \omega_1 }}  \,, \\[5mm]
{\cal J}_{2} &\defeq &\ds \frac{\pi}{\sqrt \lambda} \, J_{2} &= &\ds u_2 \, \cos^2 \omega_1 \,,
\end{array}
\label{charges DGM}
\end{equation}
where $\eK (k)$ is complete elliptic integral of the first kind, and $\eK (1) = \infty$. Then, the relation \eqref{energy-spin dgm} follows by setting $\omega_1 = \pare{\pi - p_1}/2$\,.

\bigskip
One can estimate exponential part of the finite\,-$J_1$ corrections to the leading order, only from the above information. This is because the correction term is of order $(k')^2$, while $k' \equiv \sqrt{1 - k^2}$ can also be expressed by the angular momenta.

Let us first relate $k'$ with the complete elliptic integral of the first kind $\eK(k)$. As shown in Appendix \ref{sec:expand EFZ}, $\eK(k)$ has the asymptotic form
\begin{equation}
\eK(k) = \ln \pare{\frac{4}{k'}} + {\cal O} \pare{k'^2 \ln k'}, \qquad \pare{{\rm as}\ k \to 1} .
\end{equation}
Inverting this relation, we obtain $k' = 4 \exp \cpare{-\eK(1)}$. We express a divergent constant $\eK(1)$ by angular momenta ${\cal J}_1$ and ${\cal J}_2$\,. The expressions \eqref{charges DGM} tell us
\begin{equation}
\eK (1) = \frac{1}{1 - \frac{\tan^2 \omega_1}{u_1^2}} \pare{ \frac{{\cal J}_1}{u_1} + \cos^2 \omega_1} \,,\qquad {\rm where} \quad u_1 = \frac{\sqrt{{\cal J}_2^2 + \cos^2 \omega_1}}{\cos^2 \omega_1} \,.
\end{equation}
Eliminating $u_1$ from the first equation, we get
\begin{alignat}{1}
\eK (1) &= \frac{{\cal J}_2^2 + \cos^2 \omega_1}{{\cal J}_2^2 + \cos^4 \omega_1} \, \pare{ \frac{{\cal J}_1 \cos^2 \omega_1}{\sqrt{{\cal J}_2^2 + \cos^2 \omega_1}} + \cos^2 \omega_1 },  \notag \\[1mm]
&\approx \frac{{\cal J}_2^2 + \sin^2 \frac{p_1}{2}}{{\cal J}_2^2 + \sin^4 \frac{p_1}{2}} \, \pare{ \frac{{\cal J}_1 \sin^2 \frac{p_1}{2}}{\sqrt{{\cal J}_2^2 + \sin^2 \frac{p_1}{2}}} + \sin^2 \frac{p_1}{2} } ,
\label{dyonic exp}
\end{alignat}
where we neglected higher-order corrections to the relation $2 \ssp \omega_1 = \pi - p_1 + {\cal O} \pare{k'^2}$ in the second line.

If we take the limit ${\cal J}_2 \to 0$ within this expression, we get
\begin{equation}
\eK (1) \to \frac{{\cal J}_1}{\cos \omega_1} + 1 \approx \frac{{\cal J}_1}{\sin \frac{p_1}{2}} + 1 \,,
\end{equation}
which is the single-spin result \cite{AFZ06a, AFGS07}.

\subsection{Helical strings with two spins near $\bmt{k=1}$}

For general value of \;$k$, helical strings have two finite angular momenta $J_1 \,, J_2$ and two finite winding numbers $N_1 \,, N_2$\,. Correspondingly, there are four controllable parameters $\pare{k, U, \omega_1\,, \omega_2}$. Other parameters which appear in the profile of helical strings can be expressed as functions of those four parameters. Below, we are going to investigate the precise form of these functions when $k$ is near $1$, and determine finite\,-$J_1$ correction to the energy-spin relation of dyonic giant magnons.

\begin{figure}[t]
\begin{minipage}[t]{.47\textwidth}
\centering
\resizebox{!}{50mm}{\includegraphics{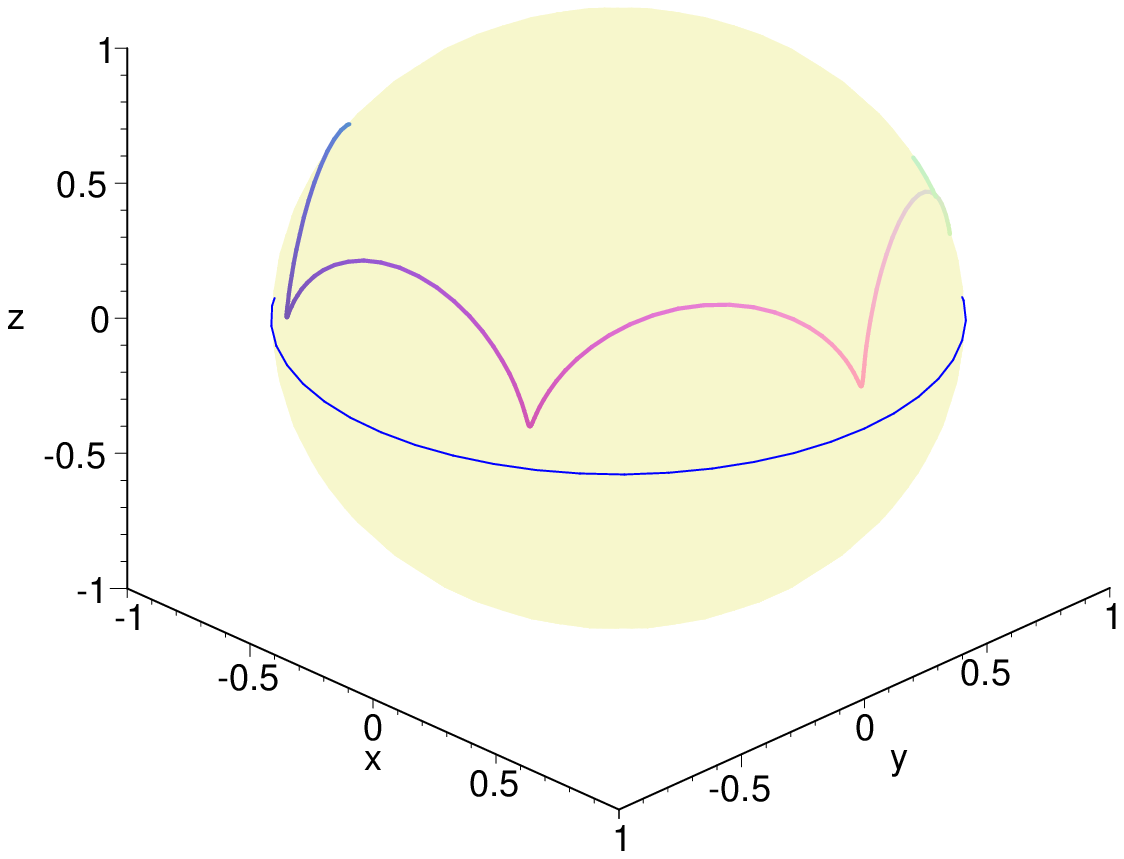}}
\end{minipage}
\hfill
\begin{minipage}[t]{.47\textwidth}
\centering
\resizebox{!}{50mm}{\includegraphics{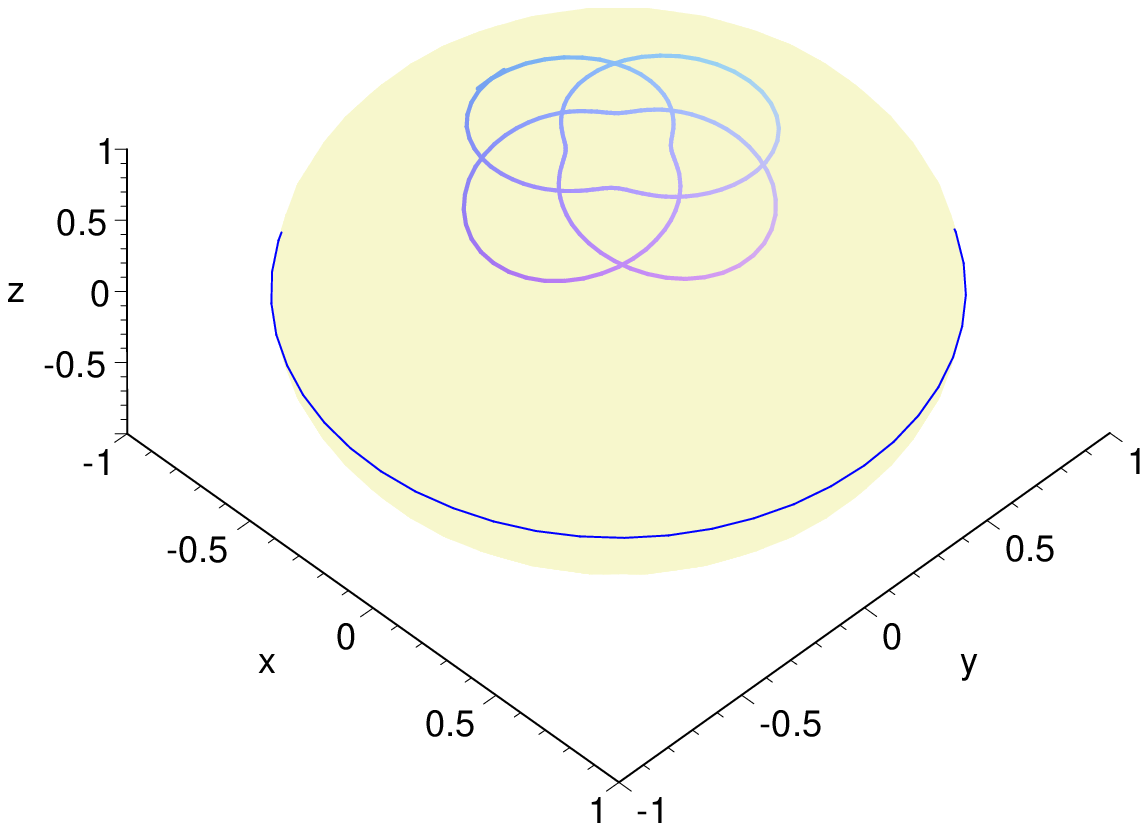}}
\end{minipage}
\caption{Left: Type $(i)$ helical spinning string solution with two spins, where the $(x,y,z)$ axes show $\pare{{\rm Re} \, \xi_1 \,, {\rm Im} \, \xi_1 \,, \abs{\xi_2}}$. Right: The same string solution with $(x,y,z) = \pare{{\rm Re} \, \xi_2 \,, {\rm Im} \, \xi_2 \,, \abs{\xi_1}}$.}
\label{fig:helical99}
\end{figure}

The profile of type $(i)$ helical string is shown in Figure \ref{fig:helical99}, and takes the following form \cite{OS06}:
\begin{align}
t &= a T + b X\,,
\label{zf0}  \\[2mm]
\xi_{1} &= C \frac{\Theta _0 (0)}{\sqrt{k} \, \Theta _0 (\iomm{1})} \frac{\Theta _1 (X   - \iomm{1})}{\Theta _0 (X )} \,
\exp \Big(  Z_0 (\iomm{1})  X+ i  u_1 T\Big)\,,
\label{zf1} \\[2mm]
\xi_{2} &= C \frac{\Theta _0 (0)}{\sqrt{k} \,\Theta _2 (\iomm{2})} \frac{\Theta _3 (X   - \iomm{2})}{\Theta _0 (X )} \, \exp \Big( Z_2 (\iomm{2} )X + i   u_2 T\Big)\,.
\label{zf2}
\end{align}
The parameters $C\,, u_1\,, u_2$ are written as
\begin{gather}
C^{-2} = \frac{\dn^2 (\iomm{2})}{k^2 \cn^2 (\iomm{2})} - \sn^2 (\iomm{1}) \,, \quad
u_1^2 = U + \dn^2 (\iomm{1}) \,,\quad
u_2^2 = U - \frac{(1 - k^2) \sn^2 (\iomm{2})}{\cn^2 (\iomm{2})} \,,
\end{gather}
and the parameters $a$ and $b$ satisfy
\begin{align}
a^2 + b^2 &= k^2 - 2 k^2 \sn^2 (\iomm{1}) - U + 2 \ssp u_2^2 \,,
\label{a,b-f1}  \\
\quad a \ssp b &= - i \, C ^2 \pare{u_1 \sn (\iomm{1}) \cn (\iomm{1}) \dn (\iomm{1}) - u_2 \, \frac{1-k^2}{k^2} \, \frac{\sn (\iomm{2}) \dn (\iomm{2})}{\cn^3 (\iomm{2})} } \,.
\label{a,b-f2}
\end{align}
The velocity $v$ is chosen so that $v \equiv b/a \le 1$.

\bigskip
All quantities given above can be expanded in powers of $k'$. Let us see the leading $k'$ corrections in turn. The normalization constant and the angular velocities become,
\begin{alignat}{1}
C &= \cos \left( \omega_1 \right) + \frac{k'^2}{4} \Big\{ \left( 1 - 2\, \cos^2 \omega_2 \right) \cos^3 \omega_1 - \cos \omega_1 + \omega_1 \sin \omega_1  \Big\} + {\cal O} (k'^4), \\[2mm]
u_1 &= {\frac {\sqrt{U \cos^2 \omega_1 + 1}}{\cos \omega_1}} - \frac{k'^2}{4} \,{\frac {\sin \omega_1 \left( \omega_1 + \sin \omega_1 \cos \omega_1 \right) }{ \cos^2 \omega_1 \sqrt {U \cos^2 \omega_1 + 1}}} + {\cal O} (k'^4), \\[1mm]
u_2 &= \sqrt {U} + \frac{k'^2}{2} \, \frac{\sin^2 \omega_2}{\sqrt{U}} + {\cal O} (k'^4).
\end{alignat}
The parameters $a, b$ and $v = b/a$ become, at the next-to-leading order,
\begin{equation}
a \approx \frac{\sqrt{U + \cos^2 \omega_1}}{\cos \omega_1} + k'^2 \, a^{(2)} \,,\quad
b \approx \tan \omega_1 + k'^2 \, b^{(2)} \,,\quad
v \approx \frac{\sin \omega_1}{\sqrt{U + \cos^2 \omega_1}} + k'^2 \, v^{(2)} \,.
\end{equation}
One can compute $a^{(2)} \,, b^{(2)}$ and $v^{(2)}$ by using the formulae shown in Appendix \ref{sec:expansion}.

We can write down the conditions for the type $(i)$ helical string \eqref{zf0}-\eqref{zf2} to be closed. If we define angular coordinates by $\varphi_{1,2} \equiv {\rm Im} \pare{ \log \, \xi_{1,2} }$, the conditions read,
\begin{alignat}{2}
\Delta \sigma \Big|_{\rm one\mbox{\tiny\,-\,}hop} &\equiv \frac{2 \pi}{n} = \frac{2 \eK (k) \sqrt{1 - v^2}}{\mu} \,,
\label{Dsigma_cl_dy} \\[2mm]
\Delta \varphi_1 \Big|_{\rm one\mbox{\tiny\,-\,}hop} &\equiv \frac{2 \pi N_1}{n} = 2 \eK (k) \pare{ -i Z_0 (\iomm{1}) - v \ssp u_1 } + (2 \ssp n'_1 + 1) \pi \,,
\label{Dphi1_cl_dy} \\[2mm]
\Delta \varphi_2 \Big|_{\rm one\mbox{\tiny\,-\,}hop} &\equiv \frac{2 \pi N_2}{n} = 2 \eK (k) \pare{ -i Z_2 (\iomm{2}) - v \ssp u_2 } + 2 \ssp n'_2 \ssp \pi \,.
\label{Dphi2_cl_dy}
\end{alignat}
As $\sigma$ runs from $0$ to $2\pi$, the string hops $n$ times in the target space, winding $N_{1}$ and $N_{2}$ times in $\varphi_1$- and $\varphi_2$-direction, respectively. One can always set $n'_{1,2} = 0$, because the shift $\omega_i \mapsto \omega_i + 2 \eK' (k)$ induces $n'_i \mapsto n'_i + 1$ with keeping the profile \eqref{zf1}, \eqref{zf2} unchanged.

The finite $J_1$ effects on the periodicity conditions can be evaluated in a similar manner. Let $p_{1,2} \equiv \Delta \varphi_{1,2}$, then the equations \eqref{Dphi1_cl_dy} and \eqref{Dphi2_cl_dy} are rewritten as, at the next-to-leading order,
\begin{alignat}{1}
p_1 &\equiv \pi - 2 \, \omega_1 + \frac{k'^2}{2} \, p_1^{(2)} + {\cal O} \pare{k'^4} \,,
\label{2spin-i:p1} \\[2mm]
p_2 &\equiv - \frac{2 \, \ell_k \sin \omega_1 \sqrt{U}}{\sqrt{U \cos^2 \omega_1 + 1}} - 2 \, \omega_2 + \frac{k'^2}{2} \, p_2^{(2)} + {\cal O} \pare{k'^4} \,,
\label{2spin-i:p2}
\end{alignat}
where $\ell_k \equiv \ln \pare{4/k'}$. We omit the exact form of $p_{1,2}^{(2)}$\,.
By inverting the relation \eqref{2spin-i:p1}, one can express $\omega_1$ in terms of $p_1$\,.
However, since $p_2$ is generally divergent as $k \to 1$, we cannot invert the relation \eqref{2spin-i:p2}. We will return to this issue in Section \ref{sec:FG interpretation}.

\bigskip
The rescaled energy ${\cal E}$ and the spins ${\cal J}_{j}$ $(j=1,2)$ were evaluated in \cite{OS06}. There we can find
\begin{align}
{\cal E} &= n \ssp a \pare{1 - v^2 }  \eK (k) \,, 
\label{zch-e} \\[2mm]
{\cal J}_1 &= \frac{n \ssp C^2 \, u_1}{{k^2 }} \cpare{ { - \eE (k) + \left( {\dn^2 (\iomm{1}) + \frac{v \ssp k^2}{{u_1 }} \ssp i \sn(\iomm{1}) \cn(\iomm{1}) \dn (\iomm{1}) } \right)\eK (k)} } \,, 
\label{zch-j1} \\[2mm]
{\cal J}_2 &= \frac{n \ssp C^2 \, u_2}{{k^2 }}\cpare{ \eE (k) + (1-k^2) \left( 
\frac{\sn^2 (\iomm{2})}{\cn^2 (\iomm{2})} - \frac{v}{u_2} \frac{i \sn(\iomm{2})\dn(\iomm{2})}{\cn^3(\iomm{2})} \right)\eK (k)} .
\label{zch-j2}
\end{align}
We may set $n=1$, since a single dyonic giant magnon corresponds to this case.
Now we expand the conserved charges in $\ell_k = \ln \pare{4/k'}$ and $k'$, and then reexpress $\omega_1$ in terms of $p_1 = \Delta \varphi_1$\,. We obtain,
\begin{alignat}{1}
{\cal E} &= \frac{\ell_k \pare{U+1} \sin \pare{\frac{p_1}{2}} }{\sqrt{U \sin^2 \pare{\frac{p_1}{2}} + 1}} + \frac{k'^2}{4} \, {\cal E}^{(2)}  + {\cal O} \pare{k'^4} ,
\label{2spin-ia:EJ2} \\[1mm]
{\cal J}_1 &= \frac{\ell_k \pare{U+1} \sin \pare{\frac{p_1}{2}} }{\sqrt{U \sin^2 \pare{\frac{p_1}{2}} + 1}} - \sqrt{U \sin^2 \pare{\frac{p_1}{2}} + 1} \, \sin \pare{\frac{p_1}{2}} + \frac{k'^2}{4} \, {\cal J}_1^{(2)} + {\cal O} \pare{k'^4} ,
\label{2spin-ib:EJ2} \\[1mm]
{\cal J}_2 &= \sqrt{U} \, \sin^2 \pare{\frac{p_1}{2}} + \frac{k'^2}{4} \, {\cal J}_2^{(2)} + {\cal O} \pare{k'^4} .
\label{2spin-ic:EJ2}
\end{alignat}
It follows that
\begin{equation}
{\cal E} - {\cal J}_1 \approx \sqrt{ {\cal J}_2^2 + \sin^2 \pare{\frac{p_1}{2}} } + \frac{k'^2}{4} \Bigg( {\cal E}^{(2)} - {\cal J}_1^{(2)} - \frac{\sqrt{U} \sin \pare{\frac{p_1}{2}}}{\sqrt{U \sin^2 \pare{\frac{p_1}{2}} + 1}} \, {\cal J}_2^{(2)} \Bigg) .
\label{2spin-i:finiteJ1}
\end{equation}
where we assumed \;$\sin \pare{p_1/2} > 0$.

The precise form of the next-to-leading terms appearing in \eqref{2spin-i:finiteJ1} can be computed with the help of formulae in Appendix \ref{sec:expansion}.
The result turns out quite simple:
\begin{equation}
{\cal E}^{(2)} - {\cal J}_1^{(2)} - \frac{\sqrt{U} \sin \pare{\frac{p_1}{2}}}{\sqrt{U \sin^2 \pare{\frac{p_1}{2}} + 1}} \, {\cal J}_2^{(2)} \ \approx \ \sin^3 \pare{\frac{p_1}{2}}  \frac{\pare{1 - 2 \cos^2 \omega_2}}{\sqrt{U \sin^2 \pare{\frac{p_1}{2}}+1}} \,.
\label{2spin-ia:NL}
\end{equation}
At this order of validity, it can also be reexpressed as
\begin{equation}
{\cal E}^{(2)} - {\cal J}_1^{(2)} - \frac{{\cal J}_2}{\sqrt{ {\cal J}_2^2 + \sin^2 \pare{\frac{p_1}{2}} }} \, {\cal J}_2^{(2)} \ \approx \ \sin^4 \pare{\frac{p_1}{2}} \frac{\pare{1 - 2 \cos^2 \omega_2}}{\sqrt{{\cal J}_2^2 + \sin^2 \pare{\frac{p_1}{2}} }} \,.
\label{2spin-ib:NL}
\end{equation}

\bigskip
For later purpose, let us introduce a new `rapidity' variable $\theta$ by
\begin{equation}
\tanh \pare{\frac{\theta}{2}} = \frac{{\cal J}_2}{\sqrt{ {\cal J}_2^2 + \sin^2 \pare{\frac{p_1}{2}} }} = \frac{\sqrt{U} \sin \pare{\frac{p_1}{2}}}{\sqrt{U \sin^{2} \pare{\frac{p_1}{2}} + 1}} + {\cal O} \pare{k'^2} \,,
\label{def-theta}
\end{equation}
then it follows
\begin{equation}
\cosh \pare{\frac{\theta}{2}} = \frac{\sqrt{ {\cal J}_2^2 + \sin^2 \pare{\frac{p_1}{2}}}}{ \sin \pare{\frac{p_1}{2}} } \approx \sqrt{U \sin^2 \pare{\frac{p_1}{2}} + 1} \,.
\end{equation}
Using this rapidity variable, \eqref{2spin-ib:NL} is rewritten as
\begin{equation}
{\cal E}^{(2)} - {\cal J}_1^{(2)} - \tanh \pare{\frac{\theta}{2}} \, {\cal J}_2^{(2)}  =  \sin^3 \pare{\frac{p_1}{2}}  \frac{\pare{1 - 2 \cos^2 \omega_2}}{\cosh \pare{\frac{\theta}{2}}} \,,
\label{2spin-i:NL2}
\end{equation}
which is the prefactor of the leading finite\,-$J_1$ correction.

\bigskip
For the exponential part, recall that $k'$ is related to $J_1$ as in \eqref{dyonic exp}:
\begin{alignat}{1}
k' &\approx  4 \, \exp \cpare{ - \frac{\sin^2 \pare{\frac{p_1}{2}}}{{\cal J}_2^2 + \sin^4 \pare{\frac{p_1}{2}}} \sqrt{{\cal J}_2^2 + \sin^2 \pare{\frac{p_1}{2}}} \pare{{\cal J}_1 + \sqrt{ {\cal J}_2^2 + \sin^2 \pare{\frac{p_1}{2}}} \, } } ,  \notag \\[2mm]
&= 4 \, \exp \cpare{ - \frac{\sin^2 \pare{\frac{p_1}{2}} \cosh^2 \pare{\frac{\theta}{2}}}{\sin^2 \pare{\frac{p_1}{2}} + \sinh^2 \pare{\frac{\theta}{2}}} \pare{ \frac{{\cal J}_1}{\sin \pare{\frac{p_1}{2}} \cosh \pare{\frac{\theta}{2}}} + 1 } } .
\label{2spin-i:NL2x}
\end{alignat}
Collecting the results \eqref{2spin-i:NL2} and \eqref{2spin-i:NL2x}, the energy-spin relation \eqref{2spin-i:finiteJ1} becomes
\begin{multline}
{\cal E} - {\cal J}_1 \approx \sqrt{ {\cal J}_2^2 + \sin^2 \pare{\frac{p_1}{2}} } \\[1mm]
- 4 \cos \pare{2 \ssp \omega_2} \, \frac{\sin^3 \pare{\frac{p_1}{2}}}{ \cosh \pare{\frac{\theta}{2}} } \,\exp \cpare{ - \frac{2 \sin^2 \pare{\frac{p_1}{2}} \cosh^2 \pare{\frac{\theta}{2}}}{\sin^2 \pare{\frac{p_1}{2}} + \sinh^2 \pare{\frac{\theta}{2}}} \pare{ \frac{{\cal J}_1}{\sin \pare{\frac{p_1}{2}} \cosh \pare{\frac{\theta}{2}}} + 1 } } .
\label{2spin-i:NL3}
\end{multline}
This is consistent with the finite\,-$J_1$ correction to giant magnons in the literature \cite{AFZ06a, AFGS07} if we set $\theta=0$ and $\cos \pare{2\ssp \omega_2} = 1$. In other words, their results are equivalent to the asymptotic behavior of single-spin type $(i)$ helical strings near $k=1$.

Single-spin type $(ii)$ helical strings corresponds to the case $\cos \pare{2\ssp \omega_2} = -1$. For two-spin case, the finite\,-$J_1$ correction is essentially same as \eqref{2spin-i:NL3}, because type $(ii)$ solution can be obtained via the operation
\begin{equation}
\omega_2 \ \mapsto \ \omega_2 + \eK'(1) = \omega_2 + \frac{\pi}{2} \,.
\end{equation}

\subsection{Finite-gap interpretation}\label{sec:FG interpretation}

Results in the last subsection revealed that the finite\,-$J_1$ correction to the energy-spin relation of dyonic giant magnons depends on the parameter $\omega_2$ that has not appeared in the $J_1 = \infty$ case.\footnote{When a two-spin helical string reduces to an array of dyonic giant magnons in $k \to 1$ limit, the dependence of $\omega_2$ naturally disappears whatever value it has.} Unfortunately we are unable to fix $\omega_2$ from the quantization condition for winding numbers, because the winding number $N_2$ becomes divergent as $k \to 1$ as we saw in \eqref{2spin-i:p2}.\footnote{Figure \ref{fig:helical99} indicates the reason for this ill-definedness; the endpoints of one-hop reach the origin $\abs{\xi_2} = 0$ as $k \to 1$.}
To clarify the situation we reconsider the r\^ole of the parameter $\omega_2$ from a finite-gap point of view, where the mode numbers are always quantized properly by construction.

\bigskip
It is well known that by exploiting the integrable structure of classical string action on \RS{3}, one can represent every classical string solution on \RS{3} (in conformal gauge) by a set of algebro-geometric data, an algebraic curve and Abelian integrals on it. These algebro-geometric data also specify what is called a Baker-Akhiezer vector. Conversely, with the help of Riemann theta functions, one can construct the Baker-Akhiezer vector such that it satisfy various constraints imposed on classical string solutions. Reconstruction of classical string solutions from the Baker-Akhiezer vector is straightforward. In this way, one can obtain a bijective map between a classical string solution and a set of algebro-geometric data \cite{DV06a, DV06b}. Algebraic curves are also useful to compare classical string solutions with the counterparts in gauge theory \cite{KMMZ04, BKS04, BKSZ05a, BKSZ05b}.

Finite-gap representation of giant magnon is first discussed in \cite{MTT06}. Further in \cite{Vicedo07}, it is shown that two-spin helical strings are equivalent to general elliptic finite-gap solutions of classical string action on \RS{3}, and that the limit $k \to 1$ corresponds to the situation in which the algebraic curve becomes singular. Written explicitly, the functions $Z_1 \,, Z_2$ of \cite{Vicedo07} correspond to $\xi_2 \,, \xi_1$ given in \eqref{zf2}, \eqref{zf1}, and the parameters $\tilde \rho_+\,, \tilde \rho_-$ of \cite{Vicedo07} correspond to $\omega_2 \,, \omega_1$\,, respectively.

The parameters $\tilde \rho_\pm$ are determined by the location of four branch points of the algebraic curve. The hermiticity of flat currents requires that the branch points should be located symmetrically with respect to the real axis. Following \cite{Vicedo07} let us write the branch points as
\begin{equation}
y^2 \defeq \pare{x - x_1} \pare{x - \bar x_1} \pare{x - x_2} \pare{x - \bar x_2}.
\end{equation}
We introduce the normalized holomorphic differential on this elliptic curve by
\begin{equation}
\omega \defeq \nu \Big/ \int_a \nu \,,\qquad \nu \defeq \frac{dx}{y} \,,
\end{equation}
where the integral over $a$ stands for the $a$-period. Then, the parameters $\tilde \rho_\pm$ are given by
\begin{equation}
i \ssp \omega_1 = i \tilde \rho_- = 2 \ssp \eK (k) \pare{ \int_{\infty^-}^{0^+} \omega - \frac{i \eK' (k)}{2 \eK (k)} } ,\qquad
i \ssp \omega_2 = i \tilde \rho_+ = 2 \ssp \eK (k) \pare{ \int_{\infty^+}^{0^+} \omega - \frac12 } ,
\label{rho_pm}
\end{equation}
with $\eK'(k) \equiv \eK(k')$. By using Riemann's bilinear identity, one can express the integral \;$\int_{\infty^\mp}^{0^+} \omega$\; in terms of the location of the branch points. The results are
\begin{equation}
\int_{\infty^\mp}^{0^+} \omega = \frac{i F \pare{ \varphi_\pm , k' }}{2 \ssp \eK (k)} \,, \qquad {\rm with} \quad
\tan \pare{\frac{\varphi_\pm}{2}} = \frac{\pare{\sqrt{\bar x_2} \pm \sqrt{x_1}} \pare{\sqrt{\bar x_1} + \sqrt{x_2}}}{\abs{x_1 - \bar x_2}} \,,
\label{int_omega}
\end{equation}
where $F \pare{\varphi, k}$ is the normal (or incomplete) elliptic integral of the first kind given in Appendix \ref{sec:definition}. From \eqref{rho_pm} and \eqref{int_omega}, we obtain the relation between the parameters $\omega_{1,2}$ of helical strings and the location of the branch points:
\begin{equation}
\omega_1 = F \pare{\varphi_+ , k'} - \eK' (k)\,, \qquad \omega_2 = F \pare{\varphi_- , k'} + i \eK (k).
\label{omega-varphi}
\end{equation}
It is shown in \cite{Vicedo07} that the right hand side of the second equation is always real. So we may redefine $\omega_2$ as
\begin{equation}
\omega _2  = 
\begin{cases}
\ {\rm Re} \cpare{ F \pare{\varphi_- , k'} } &\qquad \pare{{\rm for}\ \ k<1,\ k \to 1} , \\[1mm]
\ {\rm Re} \ds \cpare{ F \pare{\varphi_- , k'} } - \frac{\pi}{2} &\qquad \pare{{\rm for}\ \ k>1,\ k \to 1} .
\end{cases}
\label{omega2-varphi}
\end{equation}
This expression is more useful than \eqref{omega-varphi} for studying the behavior of $\omega_2$ near $k=1$.

\bigskip
Let us take the $k \to 1$ limit of the relation \eqref{omega-varphi}, which is equivalent to $x_2 \to x_1$\,. From the definition of $\varphi_\pm$ in \eqref{int_omega}, one finds
\begin{equation}
\tan \pare{\frac{\varphi_+}{2}} \to \pm \cot \pare{\frac{p}{4}} \,, \quad \tan \pare{\frac{\varphi_-}{2}} \to \mp \ssp i \,, \qquad {\rm with} \quad x_1 \equiv \exp \pare{\frac{ip-\theta}{2}} \,.
\label{varphi-limits}
\end{equation}
If we choose the upper sign in each equation, we find
\begin{equation}
\varphi_+ = - \, \frac{p}{2} + n_+ \pi \,, \qquad \varphi_- = - i \infty + r\,.
\end{equation}
with $n_+$ being an integer and $r$ a real number. Applying the formula \eqref{normal F2} to \eqref{omega-varphi} and setting $n_+ = 1$, we can reproduce the results in the previous subsection  $\omega_1 = \pare{\pi - p}/2$\,. Similarly we have $\omega_2 = r$ or $\omega_2 = r - \pi/2$Gin the latter case we may redefine $r$ to have $\omega_2 = r$.

\bigskip
To study the case $k$ is close but not equal to unity, one has to pull $x_2$ off from $x_1$\,. What matters here is that the direction in which $x_2$ is to be pulled off. If we write
\begin{equation}
x_2 = e^{i \alpha} \, x_1 \,, \qquad \alpha \equiv a + i \ssp b, \qquad {\rm with} \quad \abs{\alpha} \ll 1,
\end{equation}
then the former expressions \eqref{varphi-limits} are modified into
\begin{alignat}{1}
\tan \pare{\frac{\varphi_+}{2}} &= \pm \bpare{
\cot \pare{\frac{p}{4}} - \frac{a}{4 \sin^2 \pare{\frac{p}{4}}} } + {\cal O} \pare{\abs{\alpha}^2} ,
\label{varphi-limits2+} \\[1mm]
\tan \pare{\frac{\varphi_-}{2}} &= \mp 
\bpare{i + \frac{b}{2 \sin \pare{\frac{p}{2}}} } + {\cal O} \pare{\abs{\alpha}^2} .
\label{varphi-limits2-}
\end{alignat}
Note that the parameters $a$ and $b$ should be of order $k'$, as follows from the expression of elliptic modulus in terms of the location of branch points:
\begin{equation}
k' = \abs{\frac{x_1 - x_2}{x_1 - \bar x_2}} \approx \abs{\frac{\alpha}{2 \sin \pare{\frac{p}{2}}}} \ge \abs{\frac{\alpha}{2}} \,.
\end{equation}
Substituting these results into \eqref{omega-varphi} and \eqref{omega2-varphi}, one finds
\begin{equation}
\omega_1 = \pare{n_+ + \frac12} \pi - \frac{p}{2} + {\cal O} \pare{\abs{\alpha}} , \qquad
\omega_2 = r + {\cal O} \pare{\abs{\alpha}} .
\label{omega_limits}
\end{equation}
This result suggests that $\omega_2$ is left undetermined again in this finite-gap method.

%%%%%%%%%%%%%%%%%%%%%%
\section{Finite-Size Corrections to Magnon Boundstates}\label{sec:boundstate}

In this section, we calculate finite-size corrections to magnon boundstates by using the L\"{u}scher formula
known in quantum field theory, relating finite-size correction to the single-particle energy with the $S$-matrix of infinite-size system.
In the infinite-size limit, (dyonic) giant magnons correspond to solitons of (complex) sine-Gordon system, which are localized excitations of a two-dimensional theory.
Thus we can think of a (dyonic) giant magnon as the particle of an effective field theory, and use the L\"{u}scher formula to compute the finite-size effects of it.
More generally, such method will be applicable to string states corresponding to asymptotic spin chains \cite{Staudacher04, Beisert05}, but generic states which are not dual to asymptotic spin chains, may not be described in a simple way using particle-like picture.%
\footnote{We thank the reviewer of {\it Nuclear Physics B} for comments on this issue.}
Readers who are interested in derivation of the generalized L\"{u}scher formula in our case, please see Appendix \ref{app:Luscher}.
Here we focus ourselves on considering the $\mu$-term correction, which is given by\footnote{At the time of writing the version 5 of this paper, it is known that the correct formula is given by $\sum_b (-1)^{F_b} S_{ba}^{ba}$ rather than $\sum_b S_{ba}^{ba}$ \cite{HJL08, GSV08a}. Here we neglect this sign because fermionic terms are subleading in our computation.}
\begin{equation}
\delta \varepsilon^\mu_a = {\rm Re} \bpare{- i \sum_{Q_b>0} \left( 1 - \frac{\varepsilon'_{Q}(p)}{\varepsilon'_{Q_b} (q^1_*)} \right) e^{ - i q^1_* L} \; \mathop {\rm Res} \limits_{\tilde q = \tilde q_*} \, \sum_b S_{ba}^{ba} (\tilde q , p)},
\label{gen-mu-term0}
\end{equation}
where $p, q^1_*$ are the momenta of particles $a,b$ respectively and $Q_b$ is multiplet number of $b$.

There is possible contribution from the $F$-term. We expect that they do not contribute to the leading finite-size correction because the exponential part of the $F$-term seems different from that of the $\mu$-term, or negligibly small if $S$-matrix behaves regularly over the path of integration. We will discuss this point in Appendix \ref{app:F-term}.

\bigskip
{\bf Note:} This section is thoroughly revised in version 5 due to mistakes related to \eqref{a2 yx}.

\subsection{The $\bmt{su(2|2)^2}\ \bmt{S}$-matrix and its singularity}\label{sec:su(2|2) S-matrix}
Before applying the generalized L\"{u}scher formula to our case, let us briefly summarize some facts about the $su(2|2)^2\ S$-matrix.
Recall that elementary magnons appearing here are in the fundamental BPS representation of the $su(2|2)^2$ superconformal symmetry.

There are 16 kinds of such elementary magnons, among which scalar fields can form a part of boundstate multiplet.
The $Q$-magnon boundstate also belongs to a $16 \, Q^2$-dimensional BPS representation of $su(2|2)^2$ \cite{CDO06c, Beisert07}.
We refer to the number of magnons $Q$ as the multiplet number.

Let us first consider the scattering of two elementary magnons.
The two-body $S$-matrix has the following form:
\begin{align}
S(y,x)=S_0(y,x) [ S_{su(2|2)}(y,x) \otimes S_{su(2|2)}(y,x) ]\,,
\end{align}
where $S_0$ is the scalar factor expressed as
\begin{equation}
S_0(y,x)=\frac{y^{-}-x^{+}}{y^{+}-x^{-}}\cdot\frac{1-\frac{1}{x^{-}y^{+}}}{1-\frac{1}{x^{+}y^{-}}} \cdot \sigma^2(y,x)\,,
\end{equation}
and $S_{su(2|2)}$ is the $su(2|2)$ invariant $S$-matrix and determined only by the symmetry algebra \cite{Beisert05}. The dressing phase $\sigma^2 (y, x)$ takes the following form,
\begin{align}
\sigma^2(y,x) = \exp \Bigl[2 \ssp i \( \chi(y^-,x^-)-\chi(y^+,x^-)+\chi(y^+,x^+)-\chi(y^-,x^+) \) \Bigr] \,,
\label{the dressing sigma}
\end{align}
where $\chi(x,y)=\tilde{\chi}(x,y)-\tilde{\chi}(y,x)$, and
\begin{equation}
\tilde{\chi}(x,y) = \sum_{n=0}^\infty \frac{\tilde{\chi}^{(n)}(x,y)}{g^{n-1}}\,, \qquad
\tilde{\chi}^{(n)}(x,y) = \sum_{r=2}^\infty \sum_{s=r+1}^\infty \frac{-c_{r,s}^{(n)}}{(r-1)(s-1)x^{r-1}y^{s-1}}\,,
\label{the dressing}
\end{equation}
with the coefficients $c_{r,s}^{(n)}$ are given in \cite{BES06}.

When considering one of the two scattering bodies belongs to the $su(2)$ subsector, we just have to extract matrix elements of the form $E_1^1 \otimes E_i^j$ from the $S$-matrix of \cite{AFZ06b}. Written explicitly, they are given by
\begin{align}
S (y,x)&=S_0(y,x) \cpare{ a_1 E^1_1 \otimes E^1_1 + (a_1+a_2) E^2_2 \otimes E^1_1 + a_6 \( E^3_3 \otimes E^1_1 + E^4_4 \otimes E^1_1 \)}^2 \,,
\label{AFZ S-matrix}
\end{align}
where
\begin{align}
a_1 (y, x) &\equiv \frac{y^+ - x^-}{y^- - x^+} \, \frac{\eta (x) \eta (y)}{\tilde \eta (x) \tilde \eta (y)} \,,
\label{a1 yx} \\
a_2 (y, x) &\equiv \frac{(y^- - y^+)(x^- - x^+)(y^+ + x^-)}{(y^- - x^+)(y^- x^- - y^+ x^+)} \, \frac{\eta (x) \eta (y)}{\tilde \eta (x) \tilde \eta (y)} \,,
\label{a2 yx} \\
a_6 (y, x) &\equiv \frac{y^+ - x^+}{y^- - x^+} \, \frac{\eta (x)}{\tilde \eta (x)} \,,
\label{a6 yx}
\end{align}
The $su(2|2)$ invariant $S$-matrix does depend on the choice of frame $\eta$. For instance, if we take the {\it string} frame of \cite{AFZ06b}, we will obtain
\begin{equation}
\frac{\eta (x)}{\tilde \eta (x)} = \sqrt{\frac{x^+}{x^-}} \,, \qquad
\frac{\eta (y)}{\tilde \eta (y)} = \sqrt{\frac{y^-}{y^+}} \,.
\label{string frame}
\end{equation}
As for the {\it spin chain} frame, we obtain $\eta(x)/\tilde{\eta}(x)=\eta(y)/\tilde{\eta}(y)=1$.

If two magnons are in the same $su(2)$ sector, the corresponding $S$-matrix without the dressing phase in the spin chain frame is called BDS $S$-matrix and given by
\begin{align}
S_{\rm BDS}(y,x)=\frac{y^{-}-x^{+}}{y^{+}-x^{-}}\cdot\frac{1-\frac{1}{x^{-}y^{+}}}{1-\frac{1}{x^{+}y^{-}}} \, a_1(y,x)^2
= \frac{(y^+-x^-)(1-\frac{1}{y^+x^-})}{(y^--x^+)(1-\frac{1}{y^-x^+})} \,.
\end{align}

It is important to notice that the $S$-matrix of two {\it boundstates} factorizes into the product of the two-body $S$-matrix between elementary magnons, as the consequence of integrability.
$Q$-magnon boundstate has spectral parameters $x_k^\pm \ (k=1,\dots,Q)$, which satisfy the boundstate conditions 
\begin{equation}
x_k^{-}=x_{k-1}^{+}\;\;\;(k=2,\dots,Q).
\label{bound-cond}
\end{equation}
The magnon boundstate is thus characterized by the outermost variables
\begin{equation}
X^- \equiv x_1^- \qquad {\rm and} \qquad X^+ \equiv x_Q^+ \,.
\label{outermost X}
\end{equation}
The BDS $S$-matrix between boundstate $\{x_j^{\pm}\}$ and elementary magnon $y^{\pm}$ is given by
\begin{align}
\prod_{k=1}^Q S_{\rm BDS} (y,x_k) &= \prod_{j=1}^Q \frac{(y^+-x_k^-)(1-\frac{1}{y^+x_k^-})}{(y^--x_k^+)(1-\frac{1}{y^-x_k^+})}  \notag \\[1mm]
&= \frac{(y^{+}-X^{-})(1-\frac{1}{y^{+}X^{-}})}{(y^{-}-X^{+})(1-\frac{1}{y^{-}X^{+}})} \, \frac{(y^{-}-X^{-})(1-\frac{1}{y^{-}X^{-}})}{(y^{+}-X^{+})(1-\frac{1}{y^{+}X^{+}})} \equiv S_{\rm BDS}(y,X) \,,
\label{eq:S_BDS}
\end{align}
where we used \eqref{bound-cond} and \eqref{outermost X} \cite{CDO06b, Roiban06}.

Recall that the $su(2|2)$ invariant $S$-matrix given in \eqref{AFZ S-matrix} is also written as
\begin{equation}
S (y, x_k) = S_{\rm BDS} (y, x_k) \(\frac{\eta (x) \eta (y)}{\tilde \eta (x) \tilde \eta (y)} \)^2 \sum_{i,j=1}^4 \frac{a_i (y, x_k) a_j (y, x_k)}{a_1 (y, x_k)^2} \, (E^i_i \otimes E^1_1) \otimes (E_j^j \otimes E_1^1).
\end{equation}
Since the flavors $i$ or $j$ remain unchanged during each of the two-body scatterings, one can easily execute the product over $k$ in this expression. Thus we obtain the elementary-boundstate $S$-matrix as
\begin{equation}
S (y, X) = S_{\rm BDS} (y, X) \, \Sigma^2 (y, X) \cpare{ \sum_{b=1}^4 s_b (y, X) \, E_b {}^b \otimes E_{(1 \dots 1)} {}^{(1 \dots 1)} }^2,
\label{1,Q S-matrix}
\end{equation}
where $\Sigma (y, X)$ and $s_b (y, X)$ are given by
\begin{gather}
\Sigma (y, X) \equiv \prod_{k=1}^Q \sigma (y, x_k) \, \frac{\eta (x_k) \eta (y)}{\tilde \eta (x_k) \tilde \eta (y)} = \sigma (y, X) \, \frac{\eta (X)}{\tilde \eta (X)} \(\frac{\eta (y)}{ \tilde \eta (y)} \)^{Q},
\label{def:Sigma} \\[1mm]
s_1 (y, X) = 1 \,,\ \ 
s_2 (y, X) = \prod_{k=1}^Q \( 1 + \frac{a_2 (y, x_k)}{a_1 (y, x_k)} \), \ \ 
s_3 (y, X) = s_4 (y, X) = \prod_{k=1}^Q \frac{a_6 (y, x_k)}{a_1 (y, x_k)} \,.
\label{1,Q coeffs}
\end{gather}
Interestingly, the following formula holds\footnote{There is an identity for the spectral parameters of elementary magnons:
\begin{equation}
\frac{y^+ - x^-}{y^- - x^+} \, \( 1 - \frac{y^+ - x^+}{y^+ - x^-} \, \frac{1 - \frac{1}{y^- x^+}}{1 - \frac{1}{y^- x^-}} \) = 
\frac{(y^- - y^+)(x^- - x^+)(y^+ + x^-)}{(y^- - x^+)(y^+ x^+ - y^- x^-)} \,.
\end{equation}}
\begin{equation}
s_2 (y, X) = \frac{y^+ - X^+}{y^+ - X^-} \, \frac{1 - \frac{1}{y^- X^+}}{1 - \frac{1}{y^- X^-}} \,,\quad
s_3 (y, X) = \frac{y^+ - X^+}{y^+ - X^-} \frac{\tilde{\eta} (X)}{\eta (X)} \,,
\label{coeffs s2s3}
\end{equation}
which agree with the recent results of \cite{AF08a, BJ08}.

\bigskip
In order to compute the $\mu$-term \eqref{gen-mu-term0}, we have to evaluate the residue at poles of the $S$-matrix.
Then which poles should we pick up?
If one follows derivation of the $\mu$-term formula discussed in Appendix \ref{app:Luscher}, one finds that the following criteria need to be satisfied for a given pole to contribute to the $\mu$-term:
\begin{enumerate}
  \item The $L$-dependent exponential factor of \eqref{gen-mu-term0} damps.
  \item Gives the leading (or the largest) contribution.
  \item Comes from the $I_{abc}$-type diagram.\footnote{For classification of the Feynman diagrams, see Appendix \ref{app:Luscher}.}
\end{enumerate}
The first two criteria will be used to derive the leading exponential term \eqref{leading exponential p}, where we will consider splitting of an on-shell particle with charge $Q$ into two on-shell particles with $\pm 1$ and $Q \mp 1$.

The third criterion is related to the fact that, in quantum field theories, poles of $S$-matrix correspond to the scattering processes where intermediate particles become on-shell. For a given pole, one must be able to find a scattering process such that the on-shell condition for its intermediate states is equivalent to the pole condition of the $S$-matrix. The relation between poles of the $su(2|2)^2\ S$-matrix and scattering processes are investigated in detail in \cite{DHM07, DO07}.

\begin{figure}[t]
\begin{center}
\includegraphics[scale=0.8]{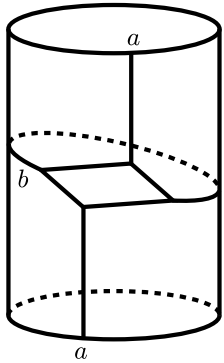}
\end{center}
\caption{The splitting process of box type.}
\label{fig:box}
\end{figure}

The third criterion states that we should pick up only the poles related to the scattering process of $I_{abc}$-type. This is so severe that various complicated processes of splitting drop out from the $\mu$-term formula. For instance, from analysis of the $S$-matrix singularity alone, the splitting process depicted in Figure \ref{fig:box} seems possible. However, this process should be classified as a $K_{ab}$-type diagram, and hence does not contribute to the $\mu$-term.

\subsection{Locating relevant poles} \label{sec:exponent}

In this section, we investigate the third criterion in detail, in order to select the poles that contribute to the $\mu$-term. As will be discussed in Appendix \ref{app:Luscher}, during the $I_{abc}$-type process an incoming particle $a$ splits into two particles $b, c$ and these two recombine into the original one after going around the worldsheet cylinder as shown in Figure \ref{fig:diagram} (Left). Importantly, the three particles $a$, $b$ and $c$ are all on-shell, and consequently for such processes to happen they must satisfy the conditions:

\medskip
3-1. Energy and momentum are conserved.

\medskip
3-2. There is a Landau-Cutkosky diagram corresponding to the process $a \to b+c$.

\bigskip
Let us first consider the conservation of energy and momentum for an on-shell splitting process $a \to b+c$. By on-shell we mean that the energy, the multiplet number, and the momentum of a (boundstate) particle are given by functions of spectral parameters $X^\pm \equiv e^{(\pm ip + \theta)/2}$ as
\begin{align}
E(X^\pm) &= \frac{g}{i} \pare{X^+ - \frac{1}{X^+} - X^- + \frac{1}{X^-} } = 4 g \cosh \Big( \frac{\theta}{2} \Big) \sin \pare{\frac{p}{2}},
\label{def:onshell E} \\[1mm]
Q(X^\pm) &= \frac{g}{i} \pare{X^+ + \frac{1}{X^+} - X^- - \frac{1}{X^-} } = 4 g \sinh \Big( \frac{\theta}{2} \Big) \sin \pare{\frac{p}{2}},
\label{def:onshell Q} \\[1mm]
p(X^\pm) &= \log \pare{\frac{X^+}{X^-}},
\end{align}
where $g = \sqrt \lambda/(4 \pi)$. The last two equations are solved as
\begin{align}
X^{\pm} \equiv e^{(\pm ip + \theta)/2} = e^{\pm ip/2} \, \frac{Q+\sqrt{Q^2+16g^2\sin^2(\frac{p}{2})}}{4g \sin(\frac{p}{2})}
= e^{\pm ip/2} \, \frac{\cQ+\sqrt{\cQ^2+\sin^2(\frac{p}{2})}}{\sin(\frac{p}{2})} \,,
\label{def:Xpm variable}
\end{align}
where $\cQ \equiv Q/(4g)$, and the parameter $\theta$ introduced above is identical to \eqref{def-theta} with ${\cal J}_2 \leftrightarrow {\cal Q}$.

Suppose the incoming particle $a$ has the multiplet number $Q = Q(X^\pm)$, the $R$-charge $r_a = Q$, and the momentum $p = p (X^\pm)$. We denote the multiplet number, and the momentum of the split particle $b$ by $Q_b$\,, $p_{b}$\,, respectively; and similarly for the other split particle $c$. Then, the conservation of energy and momentum imposes the relation:
\begin{equation}
\sqrt{Q^2+16g^2 \sin^2 \left( \frac{p}{2} \right)}=\sqrt{Q_b^2+16g^2 \sin^2 \left( \frac{p_{b}}{2} \right)}
+\sqrt{Q_c^2+16g^2 \sin^2 \left( \frac{p-p_{b}}{2} \right)}\,.
\label{eq:on-shell}
\end{equation}
We are interested in its solution that gives the smallest value of $\abs{{\rm Im} \, p_b}$, with ${\rm Im} \, p_b < 0$.
Such situation occurs when $Q_b=1$ or $Q_c=1$, and we may choose $Q_b = 1$ without loss of generality.
Further, we can constrain the multiplet number $Q_c$ by the following argument. In order that the splitting process takes place invariantly under the $su(2|2)^2$ symmetry, one should be able to contract the product of the representation of particle $b$ and that of particle $c$ with the representation of particle $a$, leaving us the singlet. In particular, if we define $\gamma \equiv Q - Q_c$, we should have $\abs{\gamma} \le 1$.%
\footnote{This argument is essentially same as in \cite{MO08}.}

Let us now solve \eqref{eq:on-shell} in the region $\abs{p_b} \ll 1$ and $Q \gg 1$. The right hand side of \eqref{eq:on-shell} can be evaluated as
\begin{align}
	{\rm R.H.S.} &\approx \sqrt{1+4g^2 p_{b}^2}+\sqrt{Q^2+16g^2 \sin^2 \left( \frac{p}{2} \right)}
	-\frac{8g^2 p_{b} \sin (\frac{p}{2}) \cos (\frac{p}{2})+\gamma Q}{\sqrt{Q^2+16g^2 \sin^2 \left( \frac{p}{2} \right)}} \,,
	\label{eq:rhs}
\end{align}
where we used $Q \gg 1$.
Inserting Eq.~(\ref{eq:rhs}) into Eq.~(\ref{eq:on-shell}), we obtain
\begin{align}
p_{b} \ \approx \ \frac{2\gamma Q\cos(\frac{p}{2}) \sin(\frac{p}{2})-\frac{i}{2g}\sqrt{(1-\gamma^2)Q^2+16g^2\sin^4(\frac{p}{2})} \sqrt{Q^2+16g^2\sin^2(\frac{p}{2})}}
	{Q^2+16g^2 \sin^4(\frac{p}{2})} \equiv q_{\ssp {\rm split}, \gamma} \,,
\label{value pb}
\end{align} 
where we choose the branch ${\rm Im} \, p_{b} < 0$.
It is easy to see that ${\rm Im} \, q_{\ssp {\rm split}, \gamma}$ reaches its minimum when $\gamma=\pm 1$,
\begin{align}
p_b=q_{\ssp {\rm split}, \pm}=\frac{\pm 2Q\cos(\frac{p}{2}) \sin(\frac{p}{2})-2i \sin^2(\frac{p}{2}) \sqrt{Q^2+16g^2\sin^2(\frac{p}{2})}}
{Q^2+16g^2 \sin^4(\frac{p}{2})}.
\label{qsplit pm}
\end{align}
From Eq.~(\ref{gen-mu-term0}), we obtain the exponential factor
\begin{align}
|e^{-i q_{\ssp {\rm split}, \pm} L}|=e^{ ({\rm Im} \, q_{\ssp {\rm split}, \pm}) \ssp L } \ \approx \ \exp \left[ -\frac{2\sin^2(\frac{p}{2})\sqrt{Q^2+16g^2\sin^2(\frac{p}{2})}}
	{Q^2+16g^2\sin^4(\frac{p}{2})} \, L \, \right].
\label{leading exponential p}
\end{align}
One can easily see that the coefficient of $L$ is same as that of $J_1$ given in \eqref{2spin-i:NL3} or \eqref{2spin-i:NL2x}.

\begin{figure}[t]
\begin{center}
\includegraphics[scale=0.8]{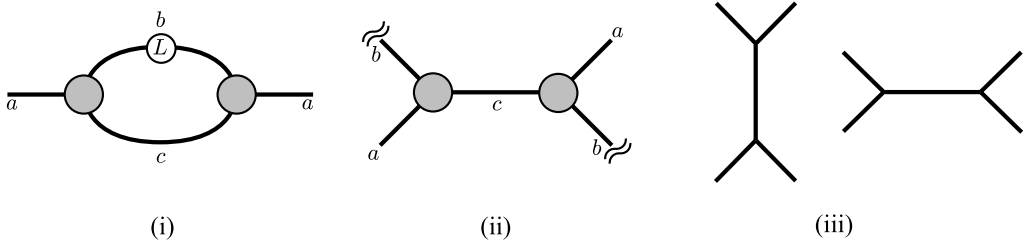}
\end{center}
\caption{(i): Self-energy diagram of $I_{abc}$-type. (ii): Diagram of $ab \to ab$ scattering made from the diagram (i). (iii): The diagram (ii) can be viewed in two ways: $s$-type diagram as shown in the left, and $t$-type diagram as shown in the right.}
\label{fig:self-scattering}

\vskip 3mm
\end{figure}

\bigskip
Next, we turn our attention to the condition 3-2. Firstly, we regard the self-energy diagrams of $I_{abc}$-type as the Landau-Cutkosky diagram of $s$- or $t$-type using the following argument (See Figure \ref{fig:self-scattering}).
If we set the particle travelling around the world, namely $b$ particle, on-shell, then self-energy diagrams of $I_{abc}$-type become equivalent to $2 \to 2$ scattering processes between particles $a$ and $b$ exchanging particle $c$, where the momenta of $a$ and $b$ remain the same after scattering. If we further put particle $c$ on-shell, this process can be expressed in terms of the Landau-Cutkosky diagram of $s$-type or $t$-type.

\begin{figure}[H]
\begin{minipage}[cbt]{0.5\linewidth}
\centering
\includegraphics[scale=0.5]{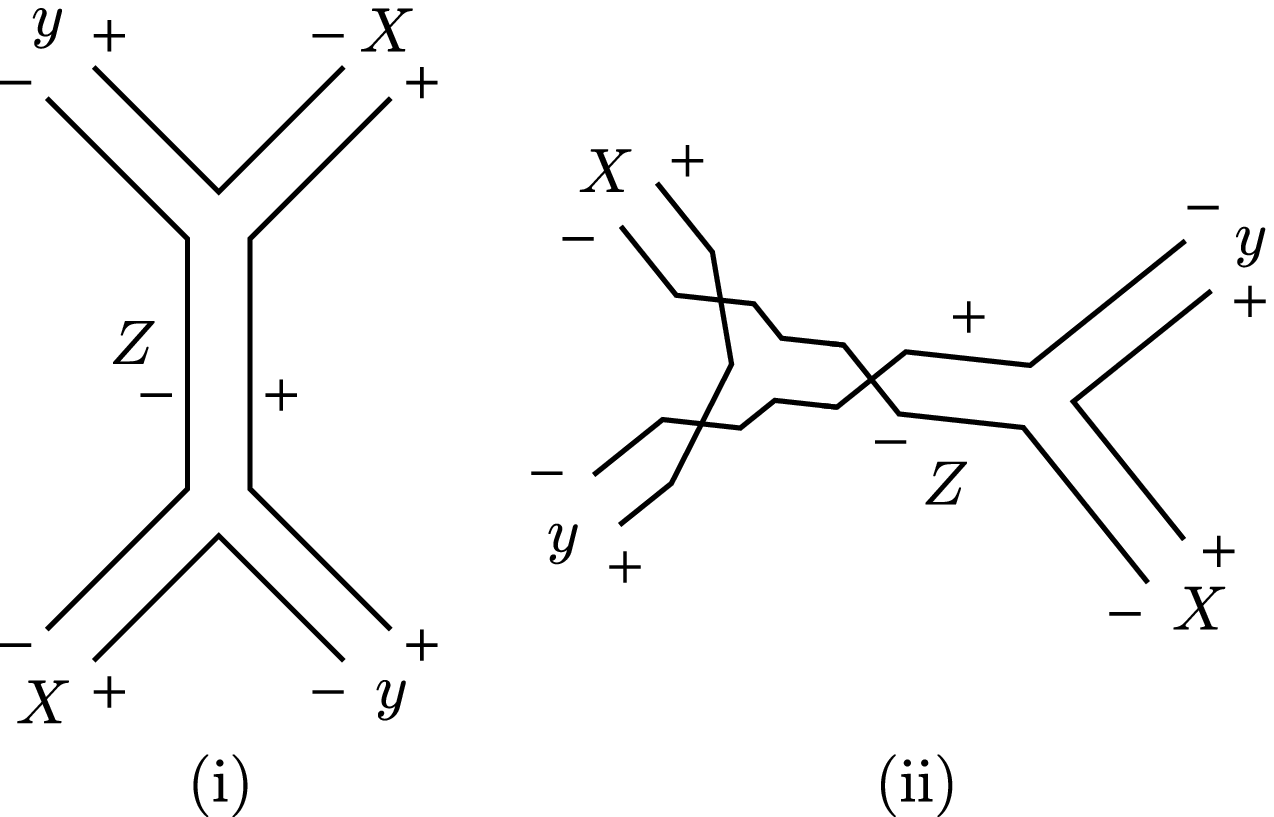}
\end{minipage}
\begin{minipage}[cbt]{0.5\linewidth}
\centering
\includegraphics[scale=0.5]{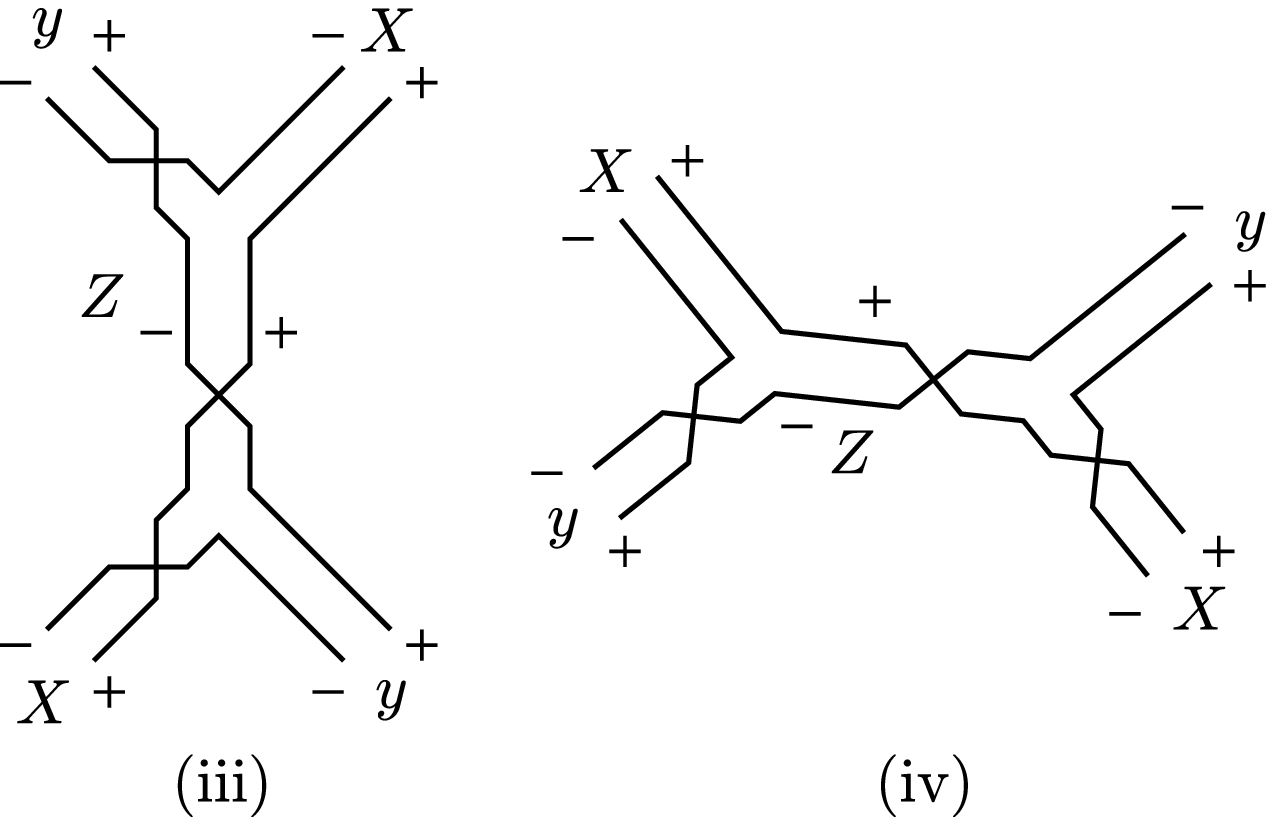}
\end{minipage}

\caption{The scattering processes which correspond to (i) $y^- = X^+$, (ii) $y^+ = X^+$, (iii) $y^- = 1/X^-$, (iv) $y^+ = 1/X^-$. We follow the convention of the diagrams in \cite{DHM07}.}
\label{fig:scattering}
\end{figure}

Secondly, for any scattering processes $a (p_a) + b(p_b) \to c (p_c) \to a (p_a) + b(p_b)$ to be kinematically allowed, it must satisfy the conservation of energy, momentum, and $R$-charge at each point of interaction. Classification of the consistent Landau-Cutkosky diagrams of $s$- or $t$-type has essentially been done in \cite{DHM07, DO07}. By following similar arguments, one can easily exhaust all consistent Landau-Cutkosky diagrams of $s$- or $t$-type. Let $X^\pm$ be the spectral parameters of the particle $a$, and $y^\pm$ be those of $b$ with $Q_b=1$, which satisfy the equation
\begin{equation}
y^{+}+\frac{1}{y^+}-y^{-}-\frac{1}{y^-}=\frac{i}{g}\,.
\label{eq:Y}
\end{equation}
Then we find four possible combinations of $\{X^\pm, y^\pm \}$ which reproduce $p_b = q_{\ssp {\rm split}, \pm}$, as listed in Table \ref{table:Iabc-scattering}. The corresponding Landau-Cutkosky diagrams of $s$- or $t$-type are shown in Figure \ref{fig:scattering}.

\bigskip
\renewcommand{\arraystretch}{1.2}

\begin{table}[t]
\centering
\caption{All possible combinations of scattering processes coming from $I_{abc}$-type diagrams which gives a damping exponential factor, namely ${\rm Im} \, p_b < 0$. Note that the crossing transformation $X^\pm \mapsto 1/X^\pm$ within this table maps the momentum with ${\rm Im} \, p_b < 0$ to the one with ${\rm Im} \, p_b > 0$. The combinations $y^- = 1/X^+$ and $y^- = X^-$ are realized as $t$-type diagram, while the ones $y^+ = 1/X^+$ and $y^+ = X^-$ are as $s$-type.}

\vskip 2mm
\begin{tabular}{c|cc|cc}
\hline
 & \multicolumn{2}{|c|}{$s$-type} & \multicolumn{2}{|c}{$t$-type} \\\hline
Pole Condition & $y^- = X^+$ & $y^- = 1/X^-$ & $y^+ = X^+$ & $y^+ = 1/X^-$ \\
In $S_{\rm BDS}$ & pole & zero & pole & zero \\
$E(Z^\pm)$ & $E(X^\pm) + E(y^\pm)$ & $E(X^\pm) + E(y^\pm)$ & $E(X^\pm) - E(y^\pm)$ & $E(X^\pm) - E(y^\pm)$ \\
$Q(Z^\pm)$ & $Q(X^\pm) + Q(y^\pm)$ & $Q(X^\pm) - Q(y^\pm)$ & $Q(X^\pm) - Q(y^\pm)$ & $Q(X^\pm) + Q(y^\pm)$ \\[1mm]
$p_b$ & $\ds \frac{-i}{2 g \sin \pare{\frac{p - i \theta}{2}}}$ & $\ds \frac{-i}{2 g \sin \pare{\frac{p + i \theta}{2}}}$ & $\ds \frac{-i}{2 g \sin \pare{\frac{p - i \theta}{2}}}$ & $\ds \frac{-i}{2 g \sin \pare{\frac{p + i \theta}{2}}}$ \\[4mm]
\hline
\end{tabular}
\label{table:Iabc-scattering}

\vskip 1.5mm
\end{table}

The processes corresponding to $y^\mp = X^+$ satisfy $p_b \approx q_{\ssp {\rm split}, +}$ and solve the condition \eqref{eq:on-shell} with $Q_c = Q-1$ at strong coupling. The ones corresponding to $y^\mp = 1/X^-$ have $p_b \approx q_{\ssp {\rm split}, -}$ and solve \eqref{eq:on-shell} with $Q_c = Q+1$.\footnote{There is no clear interpretation as such when $Q \sim \cO(1) \ll g$.} Note that this result disagrees with the classification of Table \ref{table:Iabc-scattering}. This is not contradictory, because the analyses of \cite{DHM07, DO07} are valid for arbitrary values of $g$ while ours are restricted to the case $g \to \infty$ where the solutions to the splitting condition \eqref{eq:on-shell} are degenerate.

Out of the four conditions, only the ones $y^\mp = X^+$ appear as poles of the BDS $S$-matrix \eqref{eq:S_BDS}, and the conditions $y^\mp = 1/X^-$ appear as the zeroes. The latter two actually become the poles of the full $S$-matrix because the AFS phase bring double poles at these locations. In this case, however, the spectral parameters $y^\pm$ do not lie inside the physical region $\abs{y^\pm}>1$, so we should not pick up the residues at $y^\mp = 1/X^-$.\footnote{We thank S. Frolov for a comment on physicality issue.}

In summary, we conclude that solutions to all criteria are exhausted by the two poles at $y^\mp = X^+$.

\subsection{Evaluation of residues}

We are going to evaluate the residue of each pole for the two cases $Q \sim \cO(g) \gg 1$ and $Q \sim \cO(1) \ll g$. Note that the orientation of the contour needs to be specified to fix the sign of the residue. It will turn out that the sum of two residues with the same orientation does not reproduce the results of classical string, so we will argue how the contour should be shifted to obtain the desired results.

\subsubsection{The case $\bmt{Q \sim \cO(g) \gg 1}$}\label{sec:large Q}

Let us first consider the condition $y^- = X^+$. Because $1/(y^+ - X^+) \sim \cO (g)$ around this pole, the term proportional to $s_2$ and $s_3$ in \eqref{1,Q S-matrix} are negligible at strong coupling. The residue of $S_{\rm BDS}$ is given by
\begin{equation}
\mathop {\rm Res} \limits_{\tilde q = \tilde q_*} S_{\rm BDS} (y, X) \approx \frac{(X^+ - X^-)}{(y^-)'} \, \frac{\Bigl(1 - \frac{1}{X^+ X^-} \Bigr)}{\Bigl(1 - \frac{1}{(X^+)^2} \Bigr)} \, \frac{(X^+ - X^-)}{i q^1_* X^+} \, \frac{\Bigl(1 - \frac{1}{X^+ X^-} \Bigr)}{\Bigl(1 - \frac{1}{(X^+)^2} \Bigr)} \,,
\label{y-x+ BDS}
\end{equation}
where $(y^-)'$ is the Jacobian given by \eqref{ypq Jacobian}, and we used
\begin{equation}
y^+ = \( 1+i q^1_* \) y^- + {\cal O} \pare{(q^1_*)^2} \,.
\label{Yp_approx}
\end{equation}

Next we evaluate the dressing phase. By using
\begin{equation}
\chi(y^+,X^\pm) \ \approx\ \chi(y^-, X^\pm) + i q^1_* y^- \chi_{1,0} (y^-, X^\pm)\,,
\end{equation}
we find
\begin{align}
\sigma^2(y, X) \ \approx\ \exp \Bigl[ 2 \ssp q^1_* y^- \(\chi_{1,0}(y^-, X^-)-\chi_{1,0}(y^-, X^+) \) \Bigr] \,,
\end{align}
where $\chi_{1,0}(y,x) \equiv \partial_y \chi(y,x) = \partial_x \tilde \chi(y,x) - \partial_y \tilde \chi(x, y)$. A crucial fact is that $\chi^{(n)}_{1,0}(X^{+},X^{+})$ and $\chi^{(n)}_{1,0}(X^{+},X^{-})$ are the order $1/g^{n-1}$ quantities if $Q \sim {\cal O} (\lambda^{1/2}) \gg 1$. 
The dressing phase with $n \geq 1$ does not contribute at strong coupling, which is remarkable distinction from the elementary magnon case \cite{JL07}.
Thus, it suffices to consider the contribution of $\chi^{(0)}$, namely the AFS phase \cite{AFS04}. The series \eqref{the dressing} with $c_{r,s}^{(0)} = \delta_{r+1,s}$ sums up to give
\begin{align}
\chi^{(0)}(y,x) = -g \left(\frac{1}{x}-\frac{1}{y}\right) \left( 1-(1-xy)\log \left( 1-\frac{1}{xy} \right) \right) \,.
\label{eq:chi_0}
\end{align}
It follows that
\begin{equation}
\chi^{(0)}_{1,0} (y,x) = -\frac{g}{y} \left( \frac{1}{x} + \left(y-\frac{1}{y}\right) \log \left( 1-\frac{1}{xy} \right) \right).
\end{equation}
Using this equation, the contribution of the AFS phase becomes
\begin{equation}
\sigma_{\rm AFS}^2 (y, X) \approx \exp \cpare{- \frac{2}{\Bigl( X^+ - \frac{1}{X^+} \Bigr)} \(\frac{1}{X^-} - \frac{1}{X^+}\) - 2 \ln \( \frac{1 - \frac{1}{y^- X^-}}{1 - \frac{1}{y^- X^+}} \) }.
\label{y-x+ AFS}
\end{equation}

By combining \eqref{y-x+ BDS} and \eqref{y-x+ AFS}, we find
\begin{equation}
\mathop {\rm Res} \limits_{\tilde q = \tilde q_*} S_{\rm BDS} (y, X) \; \sigma_{\rm AFS}^2 (y, X) \approx -8ig \, \frac{\sin^2 \(\frac{p}{2}\)}{\sin \(\frac{p-i\theta}{2}\)} \, \exp \cpare{- ip - \frac{\epsilon_Q (p) - Q}{2g \sin \(\frac{p-i\theta}{2}\)}} \,.
\label{y-x+ BDSAFS}
\end{equation}
To compute the $\mu$-term, one just has to multiply the prefactor
\begin{equation}
- i \pare{1 - \frac{\epsilon_Q' (p^1)}{\epsilon_1' (q_*^1)}} e^{- i q_*^1 L} = - i \, \frac{\sin \(\frac{p}{2}\) \sin \(\frac{p-i\theta}{2}\)}{\cosh\(\frac{\theta}{2}\)} \, \exp \cpare{- \frac{L}{2g \sin\(\frac{p-i\theta}{2}\)}} ,
\label{y-x+ prefactor}
\end{equation}
as well as the factor from the string frame
\begin{equation}
\frac{X^+}{X^-} \( \frac{y^-}{y^+} \)^Q \approx \exp \cpare{ip - \frac{Q}{2 g \sin \(\frac{p-i\theta}{2}\)}}.
\label{string frame factor}
\end{equation}
In total, the $\mu$-term from the pole $y^- = X^+$ is evaluated as
\begin{equation}
\delta E^\mu \Big|_{y^- = X^+} = - 8 g \, \frac{\sin^3 \(\frac{p}{2}\)}{\cosh \(\frac{\theta}{2}\)} \, \exp \cpare{- \frac{L+\epsilon_Q (p)}{2g \sin\(\frac{p-i\theta}{2}\)}} .
\label{y-x+ mu}
\end{equation}

\bigskip
Next, we study the pole $y^+ = X^+$. Now the coefficients $s_2 (y, X)$ and $s_3 (y, X)$ vanish due to \eqref{coeffs s2s3}, and only the term $s_1 (y, X)$ can contribute to the $\mu$-term. The residue of $S_{\rm BDS}$ is
\begin{equation}
\mathop {\rm Res} \limits_{\tilde q = \tilde q_*} S_{\rm BDS} (y, X) \approx \frac{(X^+ - X^-)}{- i q^1_* X^+} \, \frac{\Bigl(1 - \frac{1}{X^+ X^-} \Bigr)}{\Bigl(1 - \frac{1}{(X^+)^2} \Bigr)} \, \frac{(X^+ - X^-)}{(y^+)'} \, \frac{\Bigl(1 - \frac{1}{X^+ X^-} \Bigr)}{\Bigl(1 - \frac{1}{(X^+)^2} \Bigr)} \,.
\label{y+x+ BDS}
\end{equation}
Since $(y^+)' \approx (y^-)'$ as shown in \eqref{ypq Jacobian}, this result is just the minus of \eqref{y-x+ BDS}. The AFS phase at $y^+ = X^+$ becomes
\begin{equation}
\sigma_{\rm AFS}^2 (y, X) \approx \exp \cpare{- \frac{2}{\Bigl( X^+ - \frac{1}{X^+} \Bigr)} \(\frac{1}{X^-} - \frac{1}{X^+}\) - 2 \ln \( \frac{1 - \frac{1}{y^+ X^-}}{1 - \frac{1}{y^+ X^+}} \) },
\label{y+x+ AFS}
\end{equation}
which is equal to \eqref{y-x+ AFS}. Hence we conclude
\begin{equation}
\delta E^\mu \Big|_{y^+ = X^+} = 8 g \, \frac{\sin^3 \(\frac{p}{2}\)}{\cosh \(\frac{\theta}{2}\)} \, \exp \cpare{- \frac{L+\epsilon_Q (p)}{2g \sin\(\frac{p-i\theta}{2}\)}} .
\label{y+x+ mu}
\end{equation}
Here we neglected the orientation of contour when deriving the above results. We will discuss this issue in Section \ref{sec:compare}.

\subsubsection{The case $\bmt{Q \sim \cO(1) \ll g}$}\label{sec:small Q}

Let us now study the case $Q>1$ with $Q \ll g$, and compute the residues of \eqref{1,Q S-matrix} at $y^\pm = X^+$.
We have to evaluate the dressing phase carefully, because the terms higher order in $1/g$ contribute to the $\mu$-term, as discussed in \cite{JL07}.

Computation of the residue of the BDS $S$-matrix is straightforward, so let us focus on the dressing phase. It is useful to introduce new variables $\alpha^{ab}$ by
\begin{equation}
\frac{\alpha^{ab}}{2 g \sin \(\frac{p}{2}\)} = 1 - \frac{1}{y^a X^b}
\qquad {\rm if} \quad y^a X^b \to 1 \quad {\rm as} \quad g \to \infty.
\label{def:alpha^ab}
\end{equation}
We can neglect the higher-order terms in the dressing phase when $y^a X^b$ is not close to unity. The values of $\alpha^{ab}$ around the pole conditions are listed in Table \ref{table:alpha-ab}.

\renewcommand{\arraystretch}{1.2}
\begin{table}[tb]
\caption{List of $\alpha^{ab}$ at strong coupling, corresponding to the pole with ${\rm Im}\, q^1<0$.}

\vspace{3mm}
\centering
\begin{tabular}{c|cccc}
Pole Condition & $\alpha^{++}$ & $\alpha^{--}$ & $\alpha^{+-}$ & $\alpha^{-+}$ \\\hline
$y^- = X^+$ &  & $Q$ & $Q+1$ &  \\
$y^+ = X^+$ &  & $Q-1$ & $Q$ &  \\
\end{tabular}
\vspace{6mm}
\label{table:alpha-ab}
\end{table}

The AFS phase \cite{AFS04} can be easily computed from the following expressions:
\begin{align}
\sigma_{\rm AFS}^2 (y, X) &= \(\frac{1- \frac{1}{y^- X^-}}{1- \frac{1}{y^+ X^-}}\)^{2Q} \(\frac{1- \frac{1}{y^- X^+}}{1- \frac{1}{y^- X^-}}\)^2,
\label{AFS phase small}
\end{align}
which are derived in Appendix \ref{sec:detail}. The Hern\'andez-L\'opez phase \cite{HL06} can be computed by employing the results of \cite{JL07},
\begin{equation}
\chi^{(1)} (y^a, X^b) \approx \mp \frac{i}{2} \log \(\frac{\alpha^{ab}}{2 g \sin \(\frac{p}{2}\)}\),
\label{HL expand}
\end{equation}
where the sign ambiguity comes from the choice of a logarithmic branch. As shown in Appendix \ref{sec:detail}, the rest of the BES phase \cite{BES06} is summarized as
\begin{equation}
\sigma^2_{n \ge 2} (y, X) \approx \exp \Big[ 2 \( \alpha^{--} - \alpha^{+-} \) \Big] \(\frac{\alpha^{+-}}{\alpha^{--}}\)^{\alpha^{--} + \alpha^{+-}} ,\quad \( {\rm for} \ \ y \sim e^{ip/2} \,\),
\label{higher product}
\end{equation}
Note that $\chi^{(2m+1)} (y^a, X^b) \approx 0$. By combining the results \eqref{HL expand} and \eqref{higher product}, the higher-order dressing phase is evaluated as
\begin{alignat}{2}
\sigma^2 (y, X) &\approx - \frac{16 g^2 \sin^2 \(\frac{p}{2}\)}{Q (Q+1)} \, e^{-ip - 2} \(\frac{Q+1}{Q}\)^{\pm 1} \qquad {\rm for} \ \ y^- = X^+,
\label{y-X+ higher} \\[1mm]
\sigma^2 (y, X) &\approx
- \frac{16 g^2 \sin^2 \(\frac{p}{2}\)}{Q (Q-1)} \, e^{-ip -2} \(\frac{Q}{Q-1}\)^{\pm 1} \qquad {\rm for} \ \ y^+ = X^+.
\label{y+X+ higher}
\end{alignat}
We will choose the $+$ sign for \eqref{y-X+ higher} and the $-$ sign for \eqref{y+X+ higher} for consistency with the $Q=1$ case.\footnote{Consistency for the latter is only formal, for there is no pole at $y^+ = X^+$ when $Q=1$.}

One can calculate the remaining part of the $S$-matrix in the same manner as before. One should take care that the coefficient $s_2 (y, X)$ is non-zero for $y^-=X^+$.
The final results in string frame are summarized as
\begin{alignat}{3}
\delta E^\mu \Big|_{y^- = X^+} &= - 8 g \(1 + \frac{1}{Q} \) \sin^3 \(\frac{p}{2}\) \, \exp \cpare{- \frac{L+\epsilon_Q (p)}{2g \sin\(\frac{p}{2}\)}} ,
\label{y-x+ mu small} \\[1mm]
\delta E^\mu \Big|_{y^+ = X^+} &= + 8 g \(1 - \frac{1}{Q} \) \sin^3 \(\frac{p}{2}\) \, \exp \cpare{- \frac{L+\epsilon_Q (p)}{2g \sin\(\frac{p}{2}\)}} ,
\label{y+x+ mu small}
\end{alignat}
where $\epsilon_Q (p) \approx 4 g \sin (p/2)$.

\subsubsection{Comparison with classical string}\label{sec:compare}

Now we check if the L\"uscher $\mu$-term can reproduce the results of classical string theory, which was given in \eqref{2spin-i:NL3} as
\begin{equation}
\delta (E - J_1) = - 16 g \, \cos (2 \omega_2) \, \frac{\sin^3 \(\frac{p_1}{2}\)}{\cosh \(\frac{\theta}{2}\)} \, \exp \cpare{- \frac{\sin \(\frac{p_1}{2}\) \cosh \(\frac{\theta}{2}\)}{\sin^2 \(\frac{p_1}{2}\) + \sinh^2 \(\frac{\theta}{2}\)} \, \frac{J_1 + \epsilon_Q (p_1)}{2g} } .
\label{finite-size string}
\end{equation}

We begin with the case $Q \sim \cO(g) \gg 1$. Here, the poles $y^\pm = X^+$ are located around $\tilde q = \cot \(\frac{p-i\theta}{2}\)$, and the residues obey the relation
\begin{equation}
\delta E^\mu \Big|_{y^- = X^+} = - \delta E^\mu \Big|_{y^+ = X^+} \,.
\label{large Q residues}
\end{equation}
It suggests that the sum of $\mu$-term will vanish if we simply sum up the residues of all poles on the upper half plane. In order to obtain a nonvanishing result, for instance, we should take the difference of two residues.

\begin{figure}[t]
\centering
\includegraphics[scale=0.6]{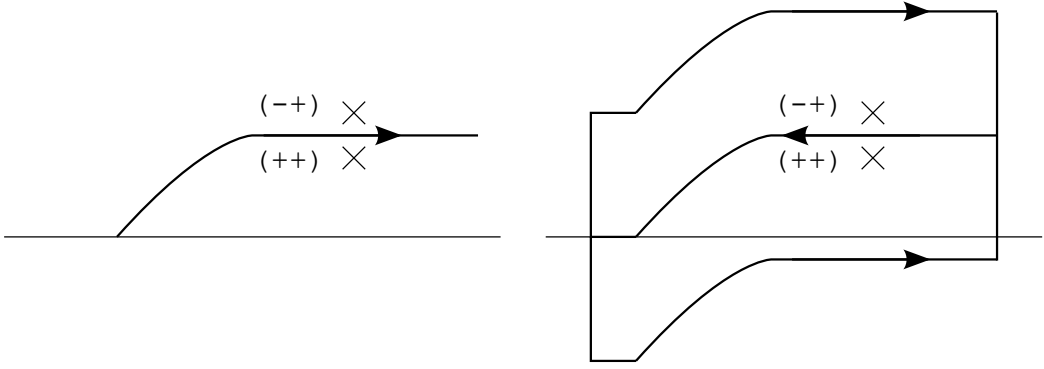}
\caption{The contour of integration in $\tilde q$ is deformed as in the left figure. By considering the difference between the $F$-term contour and the shifted contours \eqref{shift:Iabc+}, \eqref{shift:Iabc-}, we can pick up the $\mu$-term as depicted in the right figure. By $(-+), (++)$ we denote the location of poles $y^- = X^+, y^+ = X^+$, respectively.}
\label{fig:contour}
\end{figure}

We can flip the relative sign of them if we modify the contour of $\tilde q$ integration in the $F$-term formula \eqref{F-term formula} as shown in Figure \ref{fig:contour}, where $\tilde q$ is the Euclidean energy of the particle travelling around the cylinder. As discussed in Appendix \ref{app:Luscher}, we obtain the $\mu$-term from the shifts of the contour. When we set $s=1/2$ in \eqref{shift:Iabc+} and \eqref{shift:Iabc-}, we find a clockwise contour shifted upward and a counterclockwise contour shifted downward. Note that it is possible to have a clockwise contour shifted downward and a counterclockwise upward, if we choose the other branch of square root in \eqref{ypm-tq relation}, which flips the overall sign. Thus, the modified and shifted contours provide us with an additional minus sign in front of the residue at $y^+ = X^+$, giving us
\begin{equation}
\delta E^\mu \Big|_{y^- = X^+} - \delta E^\mu \Big|_{y^+ = X^+} \\
= - 16 g \ssp \cos (\alpha) \, \frac{\sin^3 \(\frac{p}{2}\)}{\cosh \(\frac{\theta}{2}\)} \, \exp \cpare{- \frac{L+\epsilon_Q (p)}{2g \sin\(\frac{p-i\theta}{2}\)}}.
\label{difference mu-term large}
\end{equation}
Since the $\mu$-term \eqref{gen-mu-term0} is given by the real part of the last expression, we obtain
\begin{equation}
\delta E^\mu = \mp 16 g \ssp \cos (\alpha) \, \frac{\sin^3 \(\frac{p}{2}\)}{\cosh \(\frac{\theta}{2}\)} \, \exp \cpare{- \frac{\sin \(\frac{p}{2}\) \cosh \(\frac{\theta}{2}\)}{\sin^2 \(\frac{p}{2}\) + \sinh^2 \(\frac{\theta}{2}\)} \, \frac{L + \epsilon_Q (p)}{2g} },
\label{total mu-term large}
\end{equation}
where
\begin{equation}
\alpha = \frac{\cos \(\frac{p}{2}\) \sinh \(\frac{\theta}{2}\)}{\sin^2 \(\frac{p}{2}\) + \sinh^2 \(\frac{\theta}{2}\)} \, \frac{L + \epsilon_Q (p)}{2g} \,,
\label{alpha mu large}
\end{equation}
for $Q \sim \cO(g) \gg 1$. This agrees with \eqref{finite-size string} upon identifying $J_1 \leftrightarrow L,\ p_1 \leftrightarrow p$ and $2 \ssp \omega_2 \leftrightarrow \alpha$.

\bigskip
Next, let us consider the case $Q \sim \cO(1) \ll g$. As shown in \eqref{exact solution r}, both poles are located on the upper half plane of the $\tilde q$ plane, namely
\begin{alignat}{2}
\tilde q = \cot \(\frac{p}{2}\) + \frac{i (Q \pm 1)}{2 g \sin^3 \(\frac{p}{2}\)} \qquad {\rm for} \quad y^\mp = X^+ .
\end{alignat}
By making the same deformation of the contour as in Figure \ref{fig:contour}, we find
\begin{equation}
\pm \( \delta E^\mu \Big|_{y^- = X^+} - \delta E^\mu \Big|_{y^+ = X^+} \) = \mp 16 g \sin^3 \(\frac{p}{2}\) \exp \cpare{ - \frac{L + \epsilon_Q (p)}{2g \sin \(\frac{p}{2}\)} },
\label{total mu-term small}
\end{equation}
for $Q \sim \cO(1) \ll g$. This result is already real, and agrees with \eqref{finite-size string} if we set $\theta = \omega_2 = 0$ and identify $J_1$ with $L$, $p_1$ with $p$.

\bigskip
Finally let us comment on computation in the spin chain frame. The result of the spin chain frame differ from that of the string frame by the factor \eqref{string frame factor}. As a consequence, the $\mu$-term for $Q \sim \cO(g)$ in \eqref{total mu-term large} turns into
\begin{equation}
\delta E^\mu = \mp 16 g \ssp \cos (\alpha_p) \, \frac{\sin^3 \(\frac{p}{2}\)}{\cosh \(\frac{\theta}{2}\)} \, \exp \cpare{- \frac{\sin \(\frac{p}{2}\) \cosh \(\frac{\theta}{2}\)}{\sin^2 \(\frac{p}{2}\) + \sinh^2 \(\frac{\theta}{2}\)} \, \frac{L - Q + \epsilon_Q (p)}{2g}},
\label{mu-term spinchain}
\end{equation}
where
\begin{equation}
\alpha_p = p + \frac{\cos \(\frac{p}{2}\) \sinh \(\frac{\theta}{2}\)}{\sin^2 \(\frac{p}{2}\) + \sinh^2 \(\frac{\theta}{2}\)} \, \frac{L - Q + \epsilon_Q (p)}{2g} \,.
\label{alpha spinchain}
\end{equation}
This expression also agrees with the result of classical string \eqref{finite-size string} if we identify $L-Q \leftrightarrow J_1,\ p_1 \leftrightarrow p$ and $2 \ssp \omega_2 \leftrightarrow \alpha_p$\,.\footnote{It appears that what we call length depends on the choice of frame.} Also, the expression \eqref{alpha spinchain} is the same as the one found in \cite{MO08}.

Thus, the $\mu$-term of the generalized L\"{u}scher formula can capture the leading finite-size (or finite angular momentum) correction to dyonic giant magnons.

%%%%%%%%%%%%%%%%%%%%%%
\section{Discussion and Conclusion}\label{sec:discussion}

In this paper, we have computed finite-size corrections to giant magnons with two angular momenta from two points of view:
\begin{itemize}
\item [(i)] Studying the asymptotic behavior of helical strings as $k \to 1$.
\item [(ii)] Slightly modifying the generalized L\"{u}scher formula and applying the $\mu$-term formula to the case in which incoming particles are boundstates.
\end{itemize}
We found that two results exactly match, which supports the validity of generalized L\"{u}scher formula for the case of boundstates.

In contrast to the work of \cite{JL07}, it turned out that when $Q \gg 1$ the leading term is only sensitive to the AFS phase in the strong coupling limit. Nevertheless, our results coincide with those in \cite{JL07} in the limit ${\cal Q} \to 0$.

\bigskip
We think the following issues are closely related to this paper and need to be clarified in the future.

It is argued in \cite{AJK05, JL07} that the exponential-type correction at strong coupling can be seen as wrapping interaction at weak coupling. We may be able to test this claim if we evaluate the generalized L\"uscher formula at weak coupling and compare it with calculation on gauge theory side, although we cannot trust the Bethe Ansatz approach at wrapping order \cite{KLRSV07, FSSZ07, KM08}.
The map from BDS spin chain \cite{BDS04} to the one-dimensional Hubbard model \cite{RSS05} might give a clue in this direction, because the Hubbard model having short-range interactions is capable of dealing with the wrapping problem.
In fact, an interesting observation has been made in \cite{JL07} that the finite-size dispersion of the Hubbard model
\begin{align}
E_{\rm Hubbard}&=\sqrt{1+16g^2 \sin^2 \left(\frac{p}{2}\right)}-\frac{2}{\sqrt{1+16g^2 \sin^2 \left(\frac{p}{2}\right)}} \, e^{-\frac{L}{2g \sin(\frac{p}{2})}}+{\cal O}\left(e^{-\frac{L}{g \sin(\frac{p}{2})}} \right),  \notag \\
& \approx 4g \sin \left( \frac{p}{2} \right)-\frac{1}{2g \sin(\frac{p}{2})} \, e^{-\frac{L}{2g \sin(\frac{p}{2})}}\,,
\label{eq:Hubbard}
\end{align}
agrees with the prediction of the generalized L\"uscher formula applied to the BDS $S$-matrix.\footnote{Strictly speaking, the Hubbard model does not agree with neither gauge nor string theory sides. For instance, the prefactor of \eqref{eq:Hubbard} does not match with the string theory result due to difference in the particle spectrum of the theory. We thank R. Janik for this remark.}

It is also interesting to study finite-size effects for (dyonic) giant magnons from matrix quantum mechanical point of view \cite{BCV05, Vazquez06, HO06, BV07}.
In the matrix quantum mechanics obtained by reducing the original ${\cal N}=4$ SYM on $\mathbb{R} \times {\rm S}^3$,
a ``string-bit", which connects two eigenvalues of background matrices forming $1/2$-BPS circular droplet, looks like a shadow of corresponding  (dyonic) giant magnon projected to 2d (or LLM) plane.
For the giant magnon with infinite-$J$, two endpoints of the string-bit localize on the edge of the circular droplet and the length of the string-bit with appropriate (normalized) radius can be interpreted as the energy of the giant magnon \cite{HM06, BCV05, Vazquez06}.

\begin{figure}[t]
\begin{center}
\includegraphics[scale=1.5]{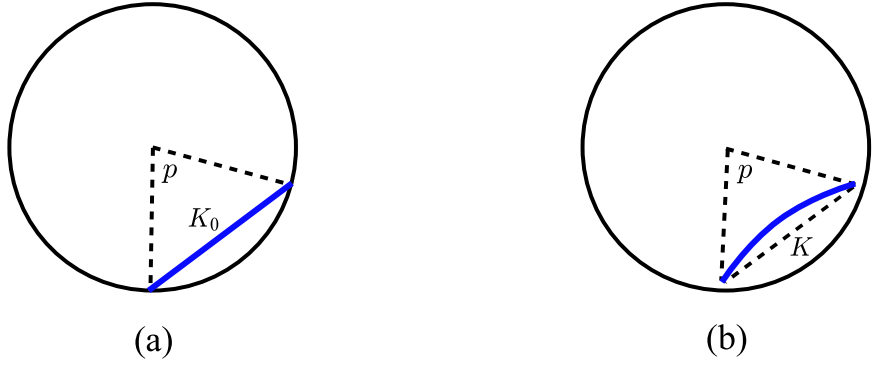}
\end{center}
\caption{The shadows of the (dyonic) giant magnons projected to 2d plane. (a) Infinite-$J$ case, (b) finite-$J$ case.
$p$ is the azimuthal angle between two endpoints, and $K$ is the distance between them (not the length of the shadow).}
\label{fig:shadows}
\end{figure}

The length of segment between endpoints of the ``finite-$J$ (dyonic) giant magnon" projected onto $\pare{{\rm Re} \, \xi_1 \,, {\rm Im} \, \xi_1}$ plane, approximately reproduces the energy-spin relation of the finite-$J$ (dyonic) giant magnon (see Figure \ref{fig:shadows}(b)). After short computation, we find
\begin{equation}
\begin{split}
\sqrt{{\cal J}_2^2 + K^2} = \sqrt{{\cal J}_2^2 + K_0^2} - \frac{K_0 \, \delta K}{\sqrt{{\cal J}_2^2 + K_0^2}} + {\cal O} \pare{ (\delta K)^2 }, \\[1mm]
{\rm with} \qquad K_0 = \sin \pare{\frac{p}{2}}, \qquad \delta K = \frac{{k'}^2}{2} \cos^2(\omega_2) \sin^3 \pare{\frac{p}{2}} \,,
\end{split}
\end{equation}
where $K = K_0 - \delta K$ is the length of {\it segment}, rather than of arc. $k'$ is the same as \eqref{2spin-i:NL2x}.
It will be interesting to investigate whether similar results can be reproduced from the matrix quantum mechanics.

The computation of one-loop quantum correction to dyonic giant magnon is also interesting. It is known that the exponential terms like $e^{-cJ}$ show up in the one-loop computation of string theory, for the case of $su(2)$ sector \cite{Sakura06} as well as of $sl(2)$ sector \cite{SZZ06}. In \cite{SZZ06}, they further discovered that quantum string Bethe Ansatz cannot reproduce such terms. We expect the generalized L\"uscher formula will also reproduce such one-loop exponential terms, as explained in Introduction.

Towards computation of the finite-size corrections exact in $L$, several approaches have been known in the theory of integrable systems, such as Thermodynamic Bethe Ansatz (TBA)\cite{Zamolodchikov90, DT96, DT97}, nonlinear integral equations (NLIE) \cite{DdV92, DdV94, DdV97, FRT98}, and functional relations among commuting transfer-matrices \cite{BLZ96}. Recently, Arutyunov and Frolov have studied TBA formulation of the finite-size system by double Wick rotation on the worksheet, and determined $S$-matrix of the ``mirror" model \cite{AF07}. Moreover, they obtained the finite-size exponential factor which is identical to the giant magnon's, by considering (two-magnon) boundstates of the mirror model. It will be very interesting to reanalyze our results from the TBA approach for multi-magnon boundstates.

\bigskip \noindent
{\it Note added}\ssp:

After submission of the paper to arXiv, we are informed of the work of J. Minahan and O. Ohlsson Sax \cite{Minahan07t, MO08} which has overlap with ours given in Section \ref{sec:DGM}.

%%%%%%%%%%%%%%%%%%%%%%%%%%%%%%%%%%%%%%%%
\subsubsection*{Acknowledgments}

We are grateful to K. Okamura and R. Janik for reading the manuscript and giving us comments. We thank K. Sakai for interesting discussion.

\appendix

%%%%%%%%%%%%%%%%%%%%%%%%%%%%%%%%%%%%%%%%%%%%%%%
\section{Definitions of Elliptic Integrals}\label{sec:definition}

For elliptic functions and the complete elliptic integrals, we follow the definitions presented in an appendix of \cite{OS06}. Below we describe the definitions of other functions and integrals which will be used in Appendix \ref{sec:expansion}.

\subsection*{Normal (or incomplete) elliptic integrals}

\begin{equation}
F (\phi, k) = \int_0^\phi \frac{d \theta}{\sqrt{ 1 - k^2 \sin^2 \theta}}
= \int_0^{\sin \phi} \frac{dt}{\sqrt{\pare{1-t^2}\pare{1-k^2 t^2}}}
\label{normal F1}
\end{equation}
is called the normal elliptic integral of the first kind.
At special values, it reduces to
\begin{equation}
F (0, k) = 0, \qquad F \pare{\frac{\pi}{2} , k} = \eK (k), \qquad
F (\phi, 0) = \phi, \qquad F (\phi, 1) = {\rm arctanh} \, \phi.
\label{normal F2}
\end{equation}

The normal elliptic integral of the first kind is related to the inverse of an elliptic function. If one regards $F (\phi, k)$ as a function of $y = \sin \phi$, then $f (y, k) \equiv F (\sin^{-1} y, k)$ obeys the differential equation
\begin{equation}
\pare{\frac{\partial f}{\partial y}}^2 = \frac{1}{\pare{1-y^2} \pare{1-k^2 y^2}} \,.
\label{F inverse1}
\end{equation}
By comparing it with
\begin{equation}
\frac{\partial \sn (z,k)}{\partial z} = \cn (z,k) \dn (z,k) = \sqrt{\pare{1 - \sn^2 (z,k)}\pare{1 - k^2 \sn^2 (z,k)}} \,,
\end{equation}
one finds
\begin{equation}
F (\phi,k) = f (y, k) = \sn^{-1} (y, k).
\label{F inverse2}
\end{equation}
The inverse of $F (\phi, k)$ as a function of $\phi$ also defines Jacobi amplitude function, by
\begin{equation}
F (\phi, k) = u \qquad \Longleftrightarrow \qquad \phi = \am (u, k).
\label{F inverse3}
\end{equation}
From \eqref{F inverse2} and \eqref{F inverse3}, it follows
\begin{equation}
\sn (u, k) = y = \sin \phi = \sin \pare{\am (u, k)} .
\end{equation}
As corollaries,
\begin{equation}
\cn (u, k) = \cos \phi, \qquad \dn (u, k) = \sqrt{1 - k^2 \sin^2 \phi \,} \qquad {\rm for} \ \ \phi = \am (u, k).
\end{equation}
We also use the notation
\begin{equation}
\eF \pare{z, k} \equiv F (\phi, k), \qquad {\rm for} \ \ \phi = \am (z, k).
\end{equation}

\bigskip
The normal (or incomplete) elliptic integral of the second kind is defined by
\begin{equation}
E (\phi, k) = \int_0^\phi d \theta \; \sqrt{1 - k^2 \sin^2 \theta}
= \int_0^{\sin \phi} dt \; \sqrt{\frac{1-k^2 t^2}{1-t^2}} \,.
\label{normal E1}
\end{equation}
We also use the notation
\begin{equation}
\eE \pare{z, k} \equiv E (\phi, k), \qquad {\rm for} \ \ \phi = \am (z, k).
\end{equation}
At special values, it reduces to
\begin{equation}
E (0, k) = 0, \qquad E \pare{\frac{\pi}{2} , k} = \eE (k), \qquad
E (\phi, 0) = \phi, \qquad E (\phi, 1) = \sin \phi.
\label{normal E2}
\end{equation}

The normal elliptic integral of the second kind is related to the integral of an elliptic function, as
\begin{equation}
E (\phi, k) = \eE \pare{z, k} = \int_0^z dw \, \dn^2 (w, k) \qquad {\rm for} \ \ \phi = \am (z,k).
\label{normal E3}
\end{equation}

\bigskip
Using \eqref{F inverse3}, and \eqref{normal E3}, one can rewrite Jacobian Zeta function as
\begin{equation}
Z_0 (z, k) = E (\phi, k) - F (\phi, k) \, \frac{\eE (k)}{\eK (k)} \qquad \pare{\phi = \am (z,k)},
\label{zeta0 2}
\end{equation}
or equivalently,
\begin{equation}
Z_0 (z, k) = \eE \pare{z, k} - z \, \frac{\eE (k)}{\eK (k)} \,.
\label{zeta0 3}
\end{equation}

Using the addition formula for Zeta functions:
\begin{equation}
Z_0 (u + v) = Z_0 (u) + Z_0 (v) - k^2 \sn(u)\sn(v)\sn(u + v),
\label{Zeta addition}
\end{equation}
one can express other Jacobi Zeta's solely by $Z_0$\,, as
\begin{alignat}{1}
Z_1 (z, k) &= Z_0 (z,k) + \frac{\cn (z,k) \dn(z,k)}{\sn (z,k)} \,, \\[1mm]
Z_2 (z, k) &= Z_0 (z,k) - \frac{\sn (z,k) \dn(z,k)}{\cn (z,k)} \,, \\[1mm]
Z_3 (z, k) &= Z_0 (z,k) - \frac{k^2 \sn (z,k) \cn(z,k)}{\dn (z,k)} \,.
\end{alignat}

\section{Expansions around $\bmt{k = 1}$}\label{sec:expansion}

\subsection{Jacobi sn, cn, and dn functions}

Jacobi sn, cn, and dn functions can be expanded in power series of ${k'}^2 \equiv 1-k^2$ around $k=1$. We want to know the expansion up to the order of $k'^4$ for later use. In an appendix of \cite{HOSV07}, the discussion is given on how to compute the expansion of Jacobi elliptic functions around $k \to 1$ analytically. Here we just cite the results:\footnote{These results can be checked also by {\tt Mathematica 6}.}
\begin{alignat}{1}
\sn (&\iom, k) \approx i \tan (\omega ) + \frac{i (1-k^2)}{4 \cos^2(\omega )} \pare{ \sin \omega \cos \omega - \omega }  \notag \\[1mm]
&+ \frac{i (1-k^2)^2}{64 \ssp \cos^3(\omega )} \, \Big( \! -9 \ssp \omega \cos \omega + \sin \omega \pare{ 4 \ssp \omega^2 + 9 - 7 \sin^2 \omega - 2 \sin^4 \omega } \Big) ,
\label{expand sn2} \\[1mm]
\cn (&\iom, k) \approx \frac{1}{\cos \omega} + \frac{1-k^2}{4 \cos^2(\omega )} \pare{ \cos \omega \sin^2 \omega - \omega \sin \omega } \notag \\[1mm]
&+ \frac{(1-k^2)^2}{64 \ssp \cos^3(\omega )} \Big( 2 \ssp \omega^2 \pare{1 + \sin^2 \omega} - \omega \sin \omega \cos \omega \pare{13 - 4 \sin^2 \omega} + 11 \sin^2 \omega \cos^2 \omega \Big) ,
\label{expand cn2}  \\[1mm]
\dn (&\iom, k) \approx \frac{1}{\cos \omega} - \frac{1-k^2}{4 \cos^2(\omega )} \pare{ \cos \omega \sin^2 \omega + \omega \sin \omega }  \notag \\[1mm]
&+ \frac{(1-k^2)^2}{64 \ssp \cos^3(\omega )} \Big( 2 \ssp \omega^2 \pare{1 + \sin^2 \omega} + \omega \sin \omega \cos \omega \pare{3 - 4 \sin^2 \omega} - 5 \sin^2 \omega \cos^2 \omega \Big) .
\label{expand dn2}
\end{alignat}

\subsection{Elliptic integrals and Jacobi Zeta functions}\label{sec:expand EFZ}

The expansion of elliptic integrals and Jacobi Zeta functions around $k=1$ is not polynomial in $k'$, because it involves $\ln k'$. Here we borrow the general results from the textbook \cite{BF71},

\paragraph{Normal elliptic integrals.}

Normal elliptic integral of the first kind behaves as
\begin{multline}
F (\phi, k) = \ln \pare{\frac{1 + \sin \phi}{\cos \phi}} - \frac{{k'}^2}{4} \cpare{\frac{\sin \phi}{\cos^2 \phi} - \ln \pare{\frac{1 + \sin \phi}{\cos \phi}}} \\[1mm]
+ \frac{3 \ssp {k'}^4}{64} \cpare{\frac{2 \sin^3 \phi}{\cos^4 \phi} - \frac{3 \sin \phi}{\cos^2 \phi} + 3 \ln \pare{\frac{1 + \sin \phi}{\cos \phi}}} + \cdots .
\label{expand nF}
\end{multline}

Normal elliptic integral of the second kind behaves as
\begin{multline}
E (\phi, k) = \sin \phi + \frac{{k'}^2}{2} \cpare{- \sin \phi + \ln \pare{\frac{1 + \sin \phi}{\cos \phi}}} \\[1mm]
- \frac{{k'}^4}{16} \cpare{\frac{\sin^3 \phi}{\cos^2 \phi} + 3 \sin \phi - 3 \ln \pare{\frac{1 + \sin \phi}{\cos \phi}}} + \cdots .
\label{expand nE}
\end{multline}

\paragraph{Complete elliptic integrals.}

Complete elliptic integral of the first kind behaves as
\begin{equation}
\eK (k) = \ln \pare{\frac{4}{k'}} + \frac{{k'}^2}{4} \cpare{\ln \pare{\frac{4}{k'}} - 1} + \frac{9 \ssp {k'}^4}{64} \cpare{\ln \pare{\frac{4}{k'}} - \frac76} + \cdots.
\label{expand cK}
\end{equation}

Complete elliptic integral of the second kind behaves as
\begin{equation}
\eE (k) = 1 + \frac{{k'}^2}{2} \cpare{ \ln \pare{\frac{4}{k'}} - \frac12} + \frac{3 {k'}^4}{16} \cpare{\ln \pare{\frac{4}{k'}} - \frac{13}{12}} + \cdots .
\label{expand cE}
\end{equation}

\bigskip
Substituting the expansion of elliptic integrals \eqref{expand nE}, \eqref{expand cK}, \eqref{expand cE} and Jacobi sn and cn functions \eqref{expand sn2}, \eqref{expand cn2}, into the expression of Jacobi Zeta \eqref{zeta0 3}, one obtains its asymptotic behavior near $k=1$:
\begin{multline}
Z_0 \pare{\iom, k} = i \tan \omega - \frac{\iom}{\ell_k} - \frac{i \ssp {k'}^2}{4} \cpare{ \frac{\omega + \sin \omega \cos \omega}{\cos^2 \omega} - \omega \pare{\frac{2}{\ell_k} - \frac{1}{\ell_k^2}} } \\[1mm]
+ \frac{i \ssp {k'}^4}{128} \cpare{ \frac{- 2 \ssp \omega \cos \omega + 2 \sin \omega \pare{ 4 \omega^2 - 5 \cos^2 \omega + 2 \cos^4 \omega }}{\cos^3 \omega} + 3 \ssp \omega \pare{\frac{4}{\ell_k} + \frac{1}{\ell_k^2} } } \\[2mm]
+ O \pare{{k'}^6} + O \pare{\frac{1}{\ell_k^3}} ,
\end{multline}
where $\ell_k \equiv \ln \pare{4/k'}$.

%%%%%%%%%%%%%%%%%%%%%%
\section{Review of the Generalized L\"{u}scher Formula}\label{app:Luscher}
In this appendix, we give a brief review on the generalized L\"{u}scher formula proposed by Janik and {\L}ukowski \cite{JL07}.
The original L\"{u}scher formula is a method to compute finite-size mass corrections from infinite-volume information of relativistic field theories \cite{Luscher83, Luscher86}.
In \cite{JL07}, this formula was generalized to the non-relativistic theory, in which an elementary particle has the dispersion relation
\begin{equation}
\varepsilon_1 (p) = \sqrt{1+16g^2 \sin^2\left(\frac{p}{2}\right)} \,,
\label{elementary disp}
\end{equation}
with $g \equiv \sqrt{\lambda}/(4\pi)$ and they reproduced the correct finite-size corrections to giant magnons.
Here we consider a little more general situation where a particle satisfies the dispersion relation of a magnon boundstate
\begin{equation}
\varepsilon_Q (p)=\sqrt{Q^2+16g^2 \sin^2\left(\frac{p}{2}\right)} \,,
\label{boundstate disp}
\end{equation}
with $Q$ an arbitrary integer. In other words, we draw a {\it single} propagator for a set of particles among whose spectral parameters satisfy the boundstate conditions $x_j^- = x_{j-1}^+$.

\begin{figure}[t]
\begin{center}
\includegraphics[scale=1.2]{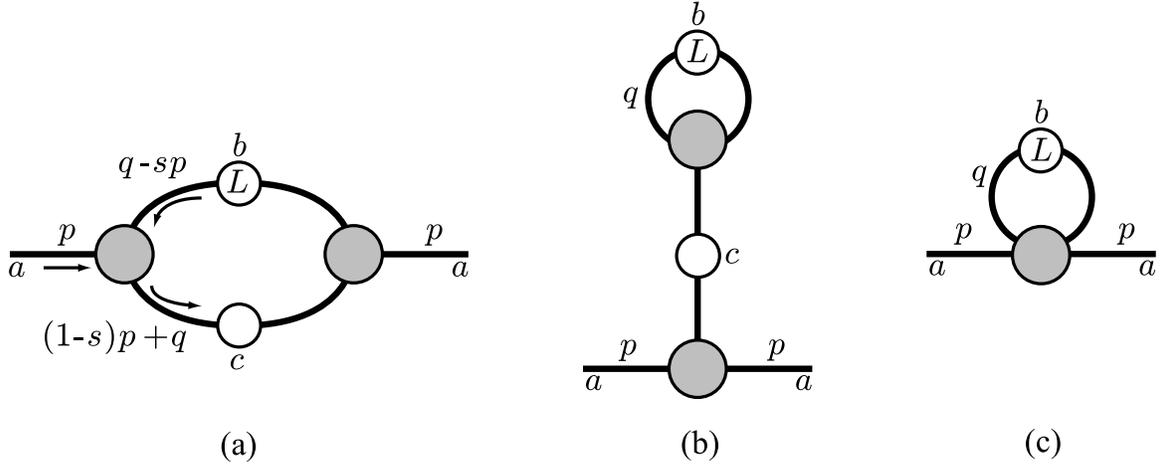}
\end{center}
\caption{Diagrams which contribute to the finite-size self-energy $\Sigma_L$\,. The propagator carrying the exponential correction is marked with $L$.}
\label{fig:IJK}
\end{figure}

\bigskip
Before deriving the generalized L\"{u}scher formula, let us make our position clearer.
We start from a two-dimensional effective Lagrangian describing the worldsheet theory in the decompactified limit.
To fix the 2-point function, we use the dispersion relations \eqref{elementary disp} and \eqref{boundstate disp} that are conjectured to all-loop orders in the 't Hooft coupling. We also assume the existence of 3- and higher point vertices, chosen so that they reproduce the conjectured two-body $S$\,-matrices.
Our treatment grounds on the following L\"uscher's argument \cite{Luscher86}. The non-perturbative nature of his formula suggests that the leading finite-size correction can be captured only by kinematics rather than dynamics, once the exact dispersion relation and $S$-matrix are known. Therefore, if we regard the magnon boundstates as a composite particle obeying the dispersion relation \eqref{boundstate disp}, we can expect generalization of L\"{u}scher formula to the dispersion relation \eqref{boundstate disp} should reproduce the correct finite-size corrections to dyonic giant magnons,

\bigskip
Now let us see derivation of the L\"{u}scher formula.
As was discussed in \cite{JL07}, the finite-size correction $\delta \varepsilon_L$ is related to the self-energy $\Sigma_L$,
\begin{align}
\delta \varepsilon_L(p)=-\frac{1}{2\varepsilon_Q(p)}\Sigma_L(p) \,.
\end{align}
There are three types of diagrams shown in Figure \ref{fig:IJK} contributing to the self-energy of particle $a$ whose charge is $Q$:
\begin{align}
(\Sigma_L)_a=\frac{1}{2}\left( \sum_{b,c}I_{abc}+\sum_{b,c}J_{abc}+\sum_{b}K_{ab} \right).
\end{align}
The term $I_{abc}$ consists of odd-point vertices, $K_{ab}$ consists of even-point vertices, and $J_{abc}$ consists of tadpole diagrams. They are given by
\begin{align}
I_{abc}&=\sum_{Q_b \ne 0} \sum_{Q_c \ne 0} \int\!\frac{d^2 q}{(2\pi)^2}\,2e^{-iq^1L}G_{b,Q_b}(q-sp)G_{c,Q_c}(q+(1-s)p) \times  \notag \\
&\hspace{3cm}	\Gamma_{abc}(-p,-q+sp,(1-s)p+q)\Gamma_{acb}(p,-(1-s)p-q,q-sp)\,,
\label{Luscher Iabc} \\
J_{abc}&=\sum_{Q_b \ne 0} \sum_{Q_c \ne 0}{}\!\!\raisebox{3mm}{\scriptsize $\prime$} \int\!\frac{d^2 q}{(2\pi)^2}\,2e^{-iq^1L}G_{b,Q_b}(q)\Gamma_{bbc}(q,-q,0) G_{c,Q_c}(0)\Gamma_{aac}(-p,p,0)\,,
\label{Luscher Jabc} \\
K_{ab}&=\sum_{Q_b \ne 0}\int\!\frac{d^2 q}{(2\pi)^2}\,2e^{-iq^1L}G_{b,Q_b}(q)\Gamma_{aabb}(p,-p,q,-q)\,,
\label{Luscher Kab} 
\end{align}
where $G$ is the (infinite-size) Green function, {\it e.g.} given by $G_{b,Q_b}(q)=((q_E^0)^2+\varepsilon_{Q_b}^2(q^1)-\Sigma(q))^{-1}$, and the $\Gamma$'s are effective 3- and 4-point vertices.
We replaced $e^{iq^1L}+e^{-iq^1L}$ with $2e^{-iq^1L}$ by an appropriate change of the loop momentum $q$, and assigned the multiplet number $Q_b,Q_c$ to the particle $b,c$ respectively, which travel around the world (see Figure \ref{fig:diagram}).
The prime over \raisebox{.5mm}{$\sum$} in \eqref{Luscher Jabc} means we sum over particles having no global $psu(2|2)^2$ charges (if such particles exist).

By shifting the contour of integration over $q^1$ in the lower imaginary-direction, we are able to neglect the integral in the limit $L\to \infty$.
We cannot however neglect the contribution from the pole of the Green functions in \eqref{Luscher Iabc}-\eqref{Luscher Kab}.
The momentum vector $(q_E^0,q^1)=(\tilde q, \tilde q^1)$ at the pole of $G_{b,Q_b}(q)$ satisfy the condition
\begin{align}
\tilde q^2+\varepsilon^2_{Q_b}(\tilde q^1)=0\,,
\end{align}
and using the dispersion relation $\varepsilon_{Q}(p)=\sqrt{Q^2+16g^2 \sin^2(\frac{p}{2})}$, we obtain
\begin{align}
\tilde q^1=-2i\arcsinh \left( \frac{\sqrt{Q_b^2+\tilde q^2}}{4g} \right).
\label{eq:q-star}
\end{align}
The integrand of $I_{abc}$ has two poles coming from $G_{b,Q_b}(q-sp)$ and $G_{c,Q_c}(q+(1-s)p)$. We denote the contribution from $G_{b,Q_b}(q-sp)$ by $I_{abc}^+$ and from $G_{c,Q_c}(q+(1-s)p)$ by $I_{abc}^-$ following \cite{JL07}. As for $I_{abc}^+$, we shift the integration variable as
\begin{equation}
q \; \mapsto \; q + sp, \qquad
G_{b,Q_b}(q-sp)G_{c,Q_c}(q+(1-s)p) \ \mapsto \ G_{b,Q_b}(q)G_{c,Q_c}(q+p),
\label{shift:Iabc+}
\end{equation}
and obtain the momentum-vector \eqref{eq:q-star}. Similarly for $I_{abc}^-$, we perform
\begin{equation}
q \; \mapsto \; q - (1-s)p, \qquad
G_{b,Q_b}(q-sp)G_{c,Q_c}(q+(1-s)p) \ \mapsto \ G_{b,Q_b}(q-p)G_{c,Q_c}(q).
\label{shift:Iabc-}
\end{equation}
Since the term $I_{abc}$ \eqref{Luscher Iabc} is symmetric under the interchange of $b$ and $c$, we obtain the same momentum-vector as in \eqref{eq:q-star} for both \eqref{shift:Iabc+} and \eqref{shift:Iabc-}.

Now using the residue of the Green function in \cite{JL07}, we can perform integration over $q^1$ and get the expression
\begin{align}
(\Sigma_L)_a=i\sum_{Q_b}\int_{-\infty}^\infty \frac{d \tilde q}{2\pi} \, \frac{e^{- i \tilde q^1 L}}{\varepsilon^2_{Q_b}(\tilde q^1)'} \; {\cal I}_a(p, \tilde q) \,,
\end{align}
where ${\cal I}_a$ is the integrand coming from the sum $I_{abc}^+ + I_{abc}^- +J_{abc}+K_{ab}$ and explicitly given by
\begin{align}
{\cal I}_a(p,q)=&\sum_{b} \sum_{c} \Bigl\{ \Gamma_{abc}(-p,-q,p+q)G_{c,Q_c}(p+q)\Gamma_{acb}(p,-p-q,q) \nonumber \\
&+\Gamma_{acb}(-p,p-q,q)G_{c,Q_c}(q-p)\Gamma_{abc}(p,-q,q-p) +\Gamma_{aabb}(p,-p,q,-q) \Bigr\} \nonumber \\[1mm]
&+\sum_b \sum_c{}\!\raisebox{3mm}{\scriptsize $\prime$} \,\Gamma_{aac}(p,-p,0)G_{c,Q_c}(0)\Gamma_{bbc}(q,-q,0),
\end{align}
where the momentum vectors $p$ and $q$ are both on-shell.
L\"uscher's remarkable observation is that the integrand ${\cal I}_a (p,q)$ is just the connected 4-point forward Green function $G_{abab}(-p,-q,p,q)$ between on-shell particles \cite{Luscher83, Luscher86, KM91}.
Furthermore, this 4-point Green function is related to the $S$-matrix element as follows:
\begin{align}
G_{abab}(-p,-q,p,q)=-4i\varepsilon_Q(p)\varepsilon_{Q_b}(q)(\varepsilon'_{Q_b}(q)-\varepsilon'_{Q}(p))(S_{ba}^{ba}(q,p)-1) \,.
\end{align}
We finally obtain the finite-size energy correction called $F$-term
\begin{align}
\delta \varepsilon_a^F(p)=-\sum_{Q_b}\int_{-\infty}^\infty\frac{d\tilde q}{2\pi}\left(1-\frac{\varepsilon'_{Q}(p)}{\varepsilon'_{Q_b}(\tilde q^1)}\right) e^{- i \tilde q^1 L}\sum_{b} (S_{ba}^{ba}(\tilde q,p)-1) \,,
\label{F-term formula}
\end{align}
where $\tilde q^1$ is given by Eq.~(\ref{eq:q-star}).

\bigskip
There is another type of the finite-size correction called $\mu$-term, which comes from the integral in $I_{abc}^\pm$.
The shifts of the integration variable made in \eqref{shift:Iabc+}, \eqref{shift:Iabc-} push the contour of integration over $q$ into the complex plane, because $q$ is Euclidean while $p$ is Minkowskian.
When we deform the contour back again onto the real axis, one may encounter new poles from the $S$-matrix.
If we denote the location of pole by $\tilde q^1 = q^1_*$, we obtain the generalized $\mu$-term formula
\begin{equation}
\delta \ssp \varepsilon^{\mu}_a(p) = -i \sum_{Q_b}\left( 1 - \frac{\varepsilon'_{Q}(p)}{\varepsilon'_{Q_b} ({q^1_*})} \right) e^{ - i q^1_* L} \; \mathop {\rm Res} \limits_{\tilde q = \tilde q_*} \, \sum_b S_{ba}^{ba}(\tilde q , p) \,.
\label{gen-mu-term}
\end{equation}

The expression \eqref{gen-mu-term} is not real-valued in general. This problem can be attributed to the replacement $\cos (i q^1 L)$ by $2e^{-iq^1L}$ to obtain the formula \eqref{Luscher Iabc}-\eqref{Luscher Kab}. If we analytically continue $q^1$ to the upper half plane, we obtain the result that is complex conjugate to \eqref{gen-mu-term}. By undoing such replacement and adding the two contributions, we obtain the real part of the above result. Consequently, the generalized $\mu$-term formula becomes
\begin{equation}
\delta \varepsilon^\mu_a = {\rm Re} \bpare{- i \sum_{Q_b>0} \left( 1 - \frac{\varepsilon'_{Q}(p)}{\varepsilon'_{Q_b} (q^1_*)} \right) e^{ - i q^1_* L} \; \mathop {\rm Res} \limits_{\tilde q = \tilde q_*} \, \sum_b S_{ba}^{ba} (\tilde q , p)},
\label{gen-mu-term-re}
\end{equation}
in place of \eqref{gen-mu-term}.

\section{$\bmt{S}$-matrix Contribution}\label{sec:detail}

\subsection{The spectral parameters and Jacobian}\label{sec:Jacobian}

The L\"uscher $F$-term formula \eqref{F-term formula} contains an integration over $\tilde q$, while the $S$-matrix is written in terms of the spectral parameters $y^\pm$. Thus in order to compute the Jacobian, we need to rewrite $y^\pm$ as functions of $\tilde q$.

The spectral parameters $y^\pm$ as functions of $q^1$ is defined by
\begin{equation}
y^\pm (q^1) = e^{\pm i q^1/2} \, \frac{Q_b + \sqrt{Q_b^2 + 16 g^2 \sin^2 \(\frac{q^1}{2}\)}}{4 g \sin \(\frac{q^1}{2}\)} \,,
\end{equation}
and the momentum $q^1$ is related to $\tilde q$ via \eqref{eq:q-star}. There are two branches of the square root, corresponding to $E(y^\pm) = \pm i \tilde q$. If we choose $E(y^\pm) = - i \tilde q$, we obtain
\begin{equation}
y^\pm (\tilde q) = \frac{\sqrt{16 g^2 + Q_b^2 + \tilde q^2} \pm \sqrt{Q_b^2 + \tilde q^2}}{4g} \, \frac{i Q_b + \tilde q}{\sqrt{Q_b^2 + \tilde q^2}} \,.
\label{ypm-tq relation}
\end{equation}
If we introduce another parameter by $\tilde q \equiv Q_b \cot (r/2)$, they translate into
\begin{equation}
y^\pm (\tilde q) = \frac{\sqrt{Q_b^2 + 16 g^2 \sin^2 \frac{r}{2}} \pm Q_b}{4 g \sin \frac{r}{2}} \, e^{ir/2} \,.
\label{ypm-r relation}
\end{equation}
Roughly speaking, the Wick rotation \eqref{eq:q-star} with $\tilde q = i q^0$ is equivalent to the transformation $(y^+, y^-) \mapsto (y^+, 1/y^-)$. 
When we set $Q_b=1$ and use \eqref{ypm-r relation}, we can solve the condition $y^\pm = X^+ $ to the next order of $1/g$ as
\begin{equation}
y^\pm = X^+ \equiv e^{(ip + \theta)/2} \qquad \Longleftrightarrow \qquad r_* \approx p-i\theta \pm \frac{i}{2g \sin\(\frac{p-i\theta}{2}\)} + \cO\(\frac{1}{g^2}\).
\label{exact solution r}
\end{equation}
Note that $\theta \approx Q/[2 g \sin(p/2)]$ if $Q \ll g$.

It is easy to compute the Jacobian between $\tilde q$ and $y^\pm$ from \eqref{ypm-tq relation}.
They read
\begin{alignat}{3}
\frac{d y^\pm (\tilde q)}{d \tilde q} \approx \frac{i}{(i - \tilde q) \sqrt{1 + \tilde q^2}} = - i \sin^2 \( \frac{r}{2} \) \, e^{ir/2} \,,
\label{ypq Jacobian}
\end{alignat}
for $g \gg 1$. Note in particular that both $(y^+)'$ and $(y^-)'$ are equal for this case.

\subsection{Dressing phase}

We will evaluate the dressing phase \eqref{the dressing sigma} for the case $Q \sim \cO(1) \ll 1$.

\paragraph{AFS phase.}

The AFS phase is given in \eqref{eq:chi_0}. Since the first term sums up to zero, the following expression is more useful:
\begin{equation}
\chi^{(0)} (y,x) = - g (y - x) \( 1 - \frac{1}{yx} \) \log \left( 1-\frac{1}{yx} \right).
\end{equation}
By using the relations
\begin{equation}
(y^+ - X^\pm)\(1-\frac{1}{y^+ X^\pm}\) = (y^- - X^\pm)\(1-\frac{1}{y^- X^\pm}\)+\frac{i}{g} \,,
\end{equation}
we find
\begin{multline}
\chi^{(0)}(y^-,X^\pm) - \chi^{(0)}(y^+,X^\pm) \\
= - g(y^- - X^\pm) \(1-\frac{1}{y^- X^\pm}\)\log\(\frac{1- \frac{1}{y^-X^\pm}}{1- \frac{1}{y^+X^\pm}}\) + i\log\(1-\frac{1}{y^+ X^\pm}\).
\label{chi1}
\end{multline}
We can relate the terms with $X^+$ to those with $X^-$ via
\begin{equation}
(y^- - X^+) \(1-\frac{1}{y^- X^+}\) = (y^- - X^-) \(1-\frac{1}{y^- X^-}\) - \frac{iQ}{g} \,.
\end{equation}
Thus we obtain
\begin{equation}
\sigma_{\rm AFS}^2 (y, X) = \(\frac{1- \frac{1}{y^- X^-}}{1- \frac{1}{y^+ X^-}}\)^{2Q} \(\frac{1- \frac{1}{y^- X^+}}{1- \frac{1}{y^- X^-}}\)^2 ,
\end{equation}
which is equal to \eqref{AFS phase small}.

\paragraph{Higher dressing phase.}

We reconsider the sum of even part of the dressing phase higher order in $1/g$. As shown in \cite{JL07}, there are contributions to the $\mu$-term from the terms $\chi^{(2m)} (y^a, X^b)$ with $y^a X^b \sim 1$ at strong coupling. If we use the variable $\alpha^{ab}$ defined by \eqref{def:alpha^ab}, the higher dressing phase can be written as
\begin{equation}
\chi^{(2m)} (\alpha^{ab}) = \pm 2i \alpha^{ab} (2m-2)! \, \frac{\zeta (2m)}{(2 \pi i \alpha^{ab})^{2m}} \,,
\label{def:schi 2m}
\end{equation}
where we take the upper sign for $y^a \sim e^{ip/2}$ and the lower sign for $y^a \sim e^{-ip/2}$. By means of Borel resummation, we can compute the summation of $\chi^{(2m)}$ over $m$ as
\begin{align}
\sum_{m=1}^\infty \chi^{(2m)} (\alpha^{ab}) &= \pm 2i \alpha^{ab} \sum_{m=1}^\infty \int_0^\infty dt \; e^{-t} \, \frac{t^{2m-2} \zeta (2m)}{(2 \pi i \alpha^{ab})^{2m}}  \notag \\
&= \pm i \int_0^\infty dt \; e^{-t} \cpare{ \frac{\alpha^{ab}}{t^2} - \frac{\coth \(\frac{t}{2 \alpha^{ab}}\)}{2t} } \,.
\end{align}
The last expression can be simplified further with the help of the following formula:\footnote{We checked this equality numerically.}
\begin{multline}
2 \int_0^\infty dt \, e^{-t} \cpare{ \frac{(\alpha^{ab} - \alpha^{cd})}{t^2} - \frac{1}{2t} \, \coth \(\frac{t}{2 \alpha^{ab}}\) + \frac{1}{2t} \, \coth \(\frac{t}{2 \alpha^{cd}}\) } \\[2mm]
= (\alpha^{ab} + \alpha^{cd}) \log \(\frac{\alpha^{ab}}{\alpha^{cd}}\) - 2 (\alpha^{ab} - \alpha^{cd}) \qquad \({\rm if} \quad \alpha^{ab} - \alpha^{cd} = \pm 1 \).
\end{multline}
The dressing phase can be computed by collecting terms with nonvanishing $\alpha^{ab}$. According to Table \ref{table:alpha-ab}, we find
\begin{equation}
\sigma^2_{n \ge 2} (y, X) \approx \exp \Big[ 2 \( \alpha^{--} - \alpha^{+-} \) \Big] \(\frac{\alpha^{+-}}{\alpha^{--}}\)^{\alpha^{--} + \alpha^{+-}} ,\quad \( {\rm for} \ \ y \sim e^{ip/2} \,\),
\end{equation}
which is \eqref{higher product}.

\section{Discussion on $\bmt{F}$-term}\label{app:F-term}

We show that $F$-term becomes negligibly small when we can avoid singularities of the $S$-matrix.

Let us first rewrite the expression for $F$-term \eqref{F-term formula} by changing integration variable. We introduce another variable $\kappa$ by
\begin{equation}
q^2 = 16 \ssp g^2 \sinh^2 \pare{\frac{\kappa}{2}} - Q_b^2 \,,\qquad \pare{q^1 = q_* \equiv - i \kappa} ,
\end{equation}
where $Q_b$ is the multiplet number of particle $b$. The $F$-term can be rewritten as
\begin{equation}
\delta \varepsilon_a^F(p) = - \sum_{Q_b \ge 1} \int_{C_{Q_b}} \frac{d \kappa}{2\pi} \, \frac{4 \ssp g^2 \sinh \kappa}{\sqrt{ 16 \ssp g^2 \sinh^2 \pare{\frac{\kappa}{2}} - Q_b^2} }
\left(1-\frac{\varepsilon'_{Q}(p)}{\varepsilon'_{Q_b}(q_*)}\right) e^{- \kappa L}\sum_{b} (S_{ba}^{ba}(q,p)-1) \,,
\label{F-term q_*}
\end{equation}
where the contour $C_{Q}$ is defined as
\begin{equation}
C_{Q} = \Big\{ \kappa \in \bb{R} \; \Big| \; \kappa \ge \kappa_{\rm cr}^{(Q)} \Big\} \,,\qquad \kappa_{\rm cr}^{(Q)} = 2 \, {\rm arcsinh} \pare{\frac{Q}{4 \ssp g}}.
\label{critical value n}
\end{equation}
Because each term within the sum at most gives the contribution $\sim e^{-\kappa_{\rm cr}^{(Q_b)} L}$, we may focus on the leading term $Q_b=1$ and rewrite it as
\begin{equation}
\delta \varepsilon_a^F(p) \Big|_{Q_b=1} \equiv - \int_{\kappa_{\rm cr}^{(1)}}^\infty d \kappa \; \frac{e^{- \kappa L}}{\sqrt{\sinh \big( \frac{\kappa}{2} \big) - \sinh \big( \frac{\kappa_{\rm cr}^{(1)}}{2} \big) }} \; f \pare{q, p}.
\label{naive F}
\end{equation}
At large $L$ the dominant contribution comes from $\kappa = \kappa_{\rm cr}^{(1)}$. If one finds singularity of $S$-matrix along the integration path, one can slightly deform the contour assuming the analyticity of integrand. Thus, if $S$-matrix behaves regularly around $\kappa = \kappa_{\rm cr}^{(1)}$, we can approximate the integral \eqref{naive F} as
\begin{equation}
\delta \varepsilon_a^F(p) \Big|_{Q_b=1}
\approx - \int_0^\infty dk \; \frac{e^{- (k + \kappa_{\rm cr}^{(1)})L}}{\sqrt k} \cdot \frac{f (-i \kappa_{\rm cr}^{(1)}, p )}{\cosh^{1/2} \big( \frac{\kappa_{\rm cr}^{(1)}}{2} \big)}
= \frac{e^{- \kappa_{\rm cr}^{(1)} L}}{\sqrt L} \cdot \frac{f (-i \kappa_{\rm cr}^{(1)}, p )}{\cosh^{1/2} \big( \frac{\kappa_{\rm cr}^{(1)}}{2} \big)} \,,
\end{equation}
which is subleading in the limit $L \to \infty$\,, because of the factor $L^{-1/2}$.

\bigskip
Singularities of the $S$-matrix appear at the position depending on the value of $X^\pm$ and $g$. And if there is a singularity at $q_* = -i \kappa_{\rm cr}^{(1)}$ which is different from single poles of the BDS $S$-matrix, the above argument will break down.
We will consider a few particular cases in which the $su(2|2)^2\ S$-matrix may possibly have singularity at $q^1 = - i \kappa_{\rm cr}^{(1)}$ in what follows.\footnote{Note that the condition $p (Y^\pm) \equiv q^1 = - i \kappa_{\rm cr}^{(1)}$ implies $E(Y^\pm) = 0$.}

Using the expression of $y^\pm$ given in Appendix \ref{sec:Jacobian}, one can find that the zeroes or the poles of the BDS $S$-matrix are found at
\begin{equation}
q^1 = \frac{- i}{2g \sin \(\frac{p \pm i \theta}{2}\)} \quad {\rm for} \ \ {\rm Im}\, q^1 < 0, \qquad
q^1 = \frac{+ i}{2g \sin \(\frac{p \pm i \theta}{2}\)} \quad {\rm for} \ \ {\rm Im}\, q^1 > 0,
\end{equation}
and they do not hit the path \eqref{critical value n} unless $p=\pi, \theta=0$. Also, by looking at \eqref{1,Q S-matrix}, one sees that the coefficients $s_2 (y, X)$ and $s_3 (y, X)$ do not bring new poles.

As discussed in \cite{DHM07, DO07}, the BHL/BES dressing phase contains an infinite number of double poles located at
\begin{equation}
X^+ + \frac{1}{X^+} - Y^- - \frac{1}{Y^-} = - \frac{im}{g} \qquad \pare{m=1,2,\ldots},
\label{double-pole location}
\end{equation}
where either one of $X^+$ or $Y^-$ must be inside the unit circle, while the other be outside. These double poles are interpreted as the kinematical constraint for the Landau-Cutkosky diagram of box type (Figure \ref{fig:box}). Below we will analytically continue $Y^\pm$ keeping particle $a$ real, $X^+ = (X^-)^*$, and study if both \eqref{double-pole location} and $q_* = -i \kappa_{\rm cr}^{(1)}$ can be solved at a particular value of $X^\pm$.

First of all, with $q_* = -i \kappa_{\rm cr}^{(1)}$ and $Q (Y^\pm)=1$, we evaluate $Y^\pm$ as,
\begin{equation}
Y^\pm = e^{\pm \frac{iq^1}{2}} \pare{ \frac{1 + \sqrt{1 + 16 g^2 \sin^2 (\frac{q^1}{2})}}{4g \sin(\frac{q^1}{2})} } \Bigg|_{q^1 = - i \kappa_{\rm cr}^{(1)}} = i \, e^{\pm \frac{\kappa_{\rm cr}^{(1)}}{2}} = i \pare{ \frac{1 \pm \sqrt{1 + 16 g^2}}{4g} },
\label{critical Ypm}
\end{equation}
showing $\abs{Y^+}>1$ and $\abs{Y^-}<1$. Plugging \eqref{critical Ypm} into \eqref{double-pole location}, we find
\begin{equation}
X^+ + \frac{1}{X^+} = - \frac{i}{2g} \pare{2m + 1} \,,
\end{equation}
which has the solutions
\begin{equation}
X^+ = i \pare{\frac{- \pare{2m+1} \pm \sqrt{\pare{2m+1}^2 + 16g^2}}{4 g}}.
\label{critical Xp}
\end{equation}
Note that we must choose the lower sign so that $X^+$ stays outside the unit circle. By using the definition $X^\pm \equiv e^{(\pm ip + \theta)/2}$ as in \eqref{def:Xpm variable}, we can identify this solution as $p = - \pi$ and $\sinh (\theta/2) = (2m+1)/4g$, which implies
\begin{equation}
X^- = - i \pare{\frac{- \pare{2m+1} - \sqrt{\pare{2m+1}^2 + 16g^2}}{4 g}}.
\label{critical Xm}
\end{equation}
However, it turns out that the spectral parameters given by \eqref{critical Xp} and \eqref{critical Xm} give rise to $Q (X^\pm) = - \pare{2m+1} <0$, which is impossible. Therefore, we conclude that there are no real values of $p$ and $\theta$ which are consistent with the double pole condition \eqref{double-pole location}, $q_* = -i \kappa_{\rm cr}^{(1)}$, $Q(Y^\pm) = 1$, and $Q(X^\pm) > 0$.

\end{document}